  \documentclass[10pt,twocolumn,twoside,letterpaper]{IEEEtran}

  \usepackage{mathtools,amsfonts,mathrsfs,amssymb,amsthm}
  \usepackage[dvips]{graphicx}
  \usepackage{float}
  \usepackage{color}
  \usepackage[plain]{algorithm}
  \usepackage{algpseudocode}
  \usepackage{siunitx}
  
  \usepackage[inline]{enumitem}
  \usepackage{framed}
  \usepackage{cite}
  \usepackage[lofdepth,lotdepth,font=footnotesize]{subfig}
  \usepackage{array}
  \usepackage{acronym}
  \usepackage[normalem]{ulem}

  \graphicspath{{Pictures/}}
  
  \restylefloat{figure} 
  
  \floatstyle{ruled} 
  \restylefloat{algorithm}

  \def\a{\mathrm{a}}
  \def\abo{\mathbf{a}}
  \def\Adm{\mathrm{Adm}}
  
  \def\b{\mathrm{b}}
  
  \def\Bbs{\boldsymbol{B}}
  \def\Bcal{\mathcal{B}}
  \def\c{\mathrm{c}}
  \def\cbo{\mathbf{c}}
  
  \def\Cfrak{\mathfrak{C}}
  \def\concat{\colon\!\!}
  \def\Const{\mathrm{Const}}
  
  \def\covbo{\mathbf{cov}}
  \def\cst{\sim}
  \def\d{\mathrm{d}}
  
  \def\Ebb{\mathbb{E}}
  \def\Exp{\mathbb{E}}
  \def\fa{\mathrm{fa}}
  
  \def\fd{\mathrm{fd}}
  \def\Fcal{\mathcal{F}}
  \def\hbo{\mathbf{h}}
  \def\Hbo{\mathbf{H}}

  \def\Ibb{\mathbb{I}}
  \def\Lcal{\mathcal{L}}
  \def\m{\mathrm{m}}
  
  \def\mbo{\mathbf{m}}
  \def\M{\mathrm{M}}
  \def\Mcal{\mathcal{M}}
  \def\md{\mathrm{md}}
  \def\Nbb{\mathbb{N}}
  \def\Nbo{\mathbf{N}}
  \def\nbs{\boldsymbol{n}}
  \def\Nbs{\boldsymbol{N}}
  
  \def\p{\mathrm{p}}
  \def\Pcal{\mathcal{P}}
  \def\Pbb{\mathbb{P}}
  
  \def\phibs{\boldsymbol{\phi}}
  \def\psibo{\mathbf{\psi}}
  \def\Rbb{\mathbb{R}}
  \def\s{\mathrm{s}}
  
  \def\st{\,|\,}
  \def\supp{\mathrm{supp}}

  \def\var{\mathrm{var}}
  \def\Var{\mathrm{Var}}
  \def\wbo{\mathbf{w}}
  \def\Xbo{\mathbf{X}}
  \def\Xcal{\mathcal{X}}
  \def\Ybo{\mathbf{Y}}

  \def\Zbo{\mathbf{Z}}

  \newcounter{subeq}
  \renewcommand{\thesubeq}{\theequation\alph{subeq}}
  \newcommand{\newsubeqblock}{\setcounter{subeq}{0}\refstepcounter{equation}}
  \newcommand{\mysubeq}{\refstepcounter{subeq}\tag{\thesubeq}}

    \newcommand{\comment}[2]{\textrm{\bf\textcolor{#1}{}}} 
    
    \newcommand{\JeremComment}[1]{\comment{blue}{#1}}
    \newcommand{\ManuComment}[1]{\comment{red}{#1}}

  \newtheorem{theoremint}{Theorem}
  \newtheorem{definitionint}{Definition}
  \newtheorem{exampleint}{Example}
  \newtheorem{remarkint}{Remark}
  \newtheorem{notationint}{Notation}
  \newtheorem{propositionint}{Proposition}
  \newtheorem{corollaryint}{Corollary}
  \newtheorem{propertyint}{Property}
  \newtheorem{lemmaint}{Lemma}
  \newtheorem*{assumptionint}{Modelling assumptions}
  \newtheorem*{approximationint}{Filtering approximations} 
  
      \newcommand*{\hdelimiter}{\vspace{0pt}}

  \newenvironment{theorem}{\hdelimiter\begin{theoremint}}{\hdelimiter\end{theoremint}}
  
  \newenvironment{example}{\hdelimiter\begin{exampleint}}{\hdelimiter\end{exampleint}}
  \newenvironment{remark}{\hdelimiter\begin{remarkint}}{\hdelimiter\end{remarkint}}

  \newenvironment{property}{\hdelimiter\begin{propertyint}}{\hdelimiter\end{propertyint}}

  \newenvironment{assumption}{\hdelimiter\begin{assumptionint}}{\hdelimiter\end{assumptionint}}

  \acrodef{cphd}[CPHD]{Cardinalized Probability Hypothesis Density}
  \acrodef{disp}[DISP]{Distinguishable and Independent Stochastic Populations}
  \acrodef{hisp}[HISP]{Hypothesised and Independent Stochastic Populations}
  \acrodef{epsrc}[EPSRC]{Engineering and Physical Sciences Research Council}
  \acrodef{fisst}[FISST]{Finite Set Statistics}
  \acrodef{gm}[GM]{Gaussian Mixture}
  \acrodef{iff}[iff]{if and only if}
  \acrodef{iid}[i.i.d.]{independently identically distributed}
  \acrodef{jpda}[JPDA]{Joint Probabilistic Density Association}
  \acrodef{lmb}[LMB]{Labeled Multi-Bernoulli}
  \acrodef{map}[MAP]{Maximum a Posteriori}
  \acrodef{mht}[MHT]{Multiple Hypothesis Tracking}
  \acrodef{phd}[PHD]{Probability Hypothesis Density}
  \acrodef{resp}[resp.]{respectively}
  \acrodef{rfs}[RFS]{Random Finite Set}
  \acrodef{smc}[SMC]{Sequential Monte Carlo}
  \acrodef{udrc}[UDRC]{University Defence Research Centre on Signal Processing}
  \acrodef{wrt}[w.r.t.]{with respect to}

  \title{Multi-object filtering with stochastic populations}
  
  \author
  {
    \IEEEauthorblockN{Emmanuel D. Delande, J\'{e}r\'{e}mie Houssineau, Daniel E. Clark}
  
    \thanks{E. D. Delande, J. Houssineau and D. E. Clark are with the School of Engineering \& Physical Sciences, Heriot-Watt University (HWU), Edinburgh (e-mails: E.D.Delande@hw.ac.uk, J.Houssineau@hw.ac.uk and D.E.Clark@hw.ac.uk).}
    
    \thanks{This work was supported by the \ac{epsrc} Platform Grant (EP/J015180/1), and the MoD \ac{udrc} Phase 2 (EP/K014227/1). J. Houssineau was sponsored by DCNS.}
  }

\begin{document}
  \maketitle
  
  \begin{abstract}
    While the design of automated knowledge-based sensor scheduling is relevant to many multi-target detection and tracking problems, tracking algorithms are rarely built for this purpose and their outputs provide little flexibility for the design of sensor policies. In this paper, we present an estimation framework for stochastic populations in the context of multi-target estimation problems. Fully probabilistic in nature, it allows for the evaluation of the population of targets through statistical moments, as well as the assessment of sensor observations through information-theoretical gain functions. We present a principled solution derived from this framework addressing challenging multi-target scenarios involving missed detections and false alarms, the filter for \acl{disp}, which propagates information on previously-detected targets as well as yet-to-be-detected targets while maintaining track continuity.
  \end{abstract}
  

  \section{Introduction} \label{sec:introduction}
    Multi-target detection and tracking problems, often constrained by limitations in sensor capabilities in realistic scenarios, are a source of inspiration and motivation for the design of automated knowledge-based sensor scheduling policies optimizing the allocation of sensing resources. Initial approaches \cite{Bar-Shalom_Y_1978, Reid_D_1979, Blackman_SS_2004} to the multi-target problem, including the \ac{mht} and \ac{jpda} techniques \cite{Blair_WD_2000}, were designed as extensions of the single-target tracking problem to the multi-target case. They rely on an intuitive approach and heuristics in order to represent the uncertainty on the number of targets in the population of interest, including ad-hoc mechanisms for track creation and deletion. Examples of exploitations of these track-based approaches in sensor scheduling problems can be found in \cite{Stuart_G_1989_1_1, Burns_PD_2004_1_1}, or in \cite[chap. 14]{Blackman_SS_1999}. The more recent \ac{fisst} framework \cite{Mahler_RPS_2007_3}, on the other hand, is fully probabilistic in nature. It allows for the principled derivation of filtering solutions such as the \ac{phd} filter \cite{Mahler_RPS_2003}, or the \ac{lmb} \cite{Vo_BT_2013} filter, and paved the way for the construction of principled sensor scheduling policies fully integrated to the multi-target estimation problem \cite{Mahler_RPS_2003_2, Mahler_RPS_2007_2, Ristic_B_2011, Andrecki_M_2015_1_2, Beard_M_2015_1_1}.
    
    In this paper, we present an estimation framework for stochastic populations \cite{Houssineau_J_2015_1_2} as an original approach to multi-target estimation problems, and a principled filtering solution derived from this framework, the filter for \ac{disp}. Some of the key features of this approach highlighted in this paper are as follows:\newline
    
    1. \emph{Stochastic populations} represent uncertainty on the composition of the population of targets \emph{and} on the targets' state. Two types of \emph{statistical moments} are presented for a stochastic population, in order to evaluate the composition of the population or the state of its members. The latter extend the regional statistics for \ac{fisst}-based filters \cite{Delande_E_2014_4}, exploited in \cite{Andrecki_M_2015_1_2} for a sensor scheduling problem.\newline     
    \indent 2. The estimation framework uses a common probabilistic description for a subpopulation of targets \emph{indistinguishable} from each other for the purpose of estimation (e.g. the targets that appeared at the same time step and have not been detected yet), so that unnecessary permutations are avoided during data association steps with observations collected from the sensor.\newline  
    \indent 3. The \ac{disp} filter propagates information on \emph{previously-detected} targets, for which track continuity is maintained through their past observations, and on \emph{yet-to-be-detected} as well, based on the operator's  knowledge about the targets entering the surveillance scene.\newline 
    \indent 4. Principled information-theoretical gain functions are derived from the output of the \ac{disp} filter, focusing on specific targets and/or specific regions of the surveillance scene, in order to quantify the information gain provided by the observations collected from the sensor system. It extends the information gain functions presented in \cite{Delande_E_2014_3_3}.\newline
    
    The paper is organized as follows. Section~\ref{sec:notation} gathers all the notations used throughout the paper. Section~\ref{sec:stochastic} presents the estimation framework for stochastic populations, and its salient features in the context of multi-target estimation problems. Section~\ref{sec:model} then presents a typical multi-target surveillance activity, and the modelling assumptions leading to the \ac{disp} filter. Section~\ref{sec:filter_representation} then shapes the modelling assumptions into the stochastic population propagated by the \ac{disp} filter, and Section~\ref{sec:filter_prediction} and Section~\ref{sec:filter_update} describe the time prediction and data update steps of the \ac{disp} filter, respectively. Section~\ref{sec:exploitation} discusses the exploitation of the \ac{disp} filter to assess the population of targets, and introduces a principled approximation of the filter with reduced complexity. Section~\ref{sec:conclusion} concludes, followed by additional material and proofs in the appendices.
    
  \section{General notations} \label{sec:notation}
%
    \subsection{Measure theory} \label{subsec:notation_measure}
      We assume that random variables are defined on some probability space $(\Omega,\Fcal,\Pbb)$ and denote $\Ebb$ the expectation \ac{wrt} $\Pbb$. We denote by $\Mcal(E)$ (\acs{resp} $\Pcal(E)$) the set of finite positive measures (\acs{resp} probability measures) on a given measurable space $(E, \Bcal(E))$, where $\Bcal(E)$ is the Borel $\sigma$-algebra on $E$.

      We shall also use the notation $\d x$, defined for any point $x \in E$, for some infinitesimal Borel set in $\Bcal(E)$ containing $x$. In the course of the paper, we shall consider augmented spaces of the form $\bar{\Xbo} = \{\psi\} \cup \Xbo$. The notation $\d x$ is then to be understood as the Borel set $\{\psi\}$ if $x = \psi$, or as some infinitesimal Borel set in $\Bcal(\Xbo)$ containing $x$, if $x \in \Xbo$.
        
      Additionally, the Banach space of all bounded and measurable functions on $E$ equipped with the uniform norm will be denoted $\Lcal^{\infty}(E)$, and we write $\mu(f) = \int \mu(\d x) f(x)$ for any $\mu \in \Mcal(E)$ and any function $f \in \Lcal^{\infty}(E)$. For any $\mu, \mu' \in \Mcal(E)$, we shall use the notation $\mu(\d x) = \mu'(\d x)$ when $\mu(f) = \mu'(f)$ for any function $f \in \Lcal^{\infty}(E)$. 
      
      Finally, for some subset $B \in \Bcal(E)$, the indicator function $1_B$ is the function on $E$ defined by
      \begin{equation} \label{eq:indicator_function}
	1_B(x) =
	\begin{dcases}
	  1, & x \in B,
	  \\
	  0, & x \notin B,
	\end{dcases}
      \end{equation}
      and the scalar $\mu(1_B)$ will sometimes be denoted $\mu(B)$, both notations being used interchangeably in the rest of the paper.  
    
    \subsection{Time prediction} \label{subsec:notation_markov_kernel}
      A transition kernel $Q : E \times \Bcal(E') \to \Rbb$ from a measurable space $E$ into a measurable space $E'$ verifies
      \begin{enumerate*}[label=\roman*)]
	\item $x \mapsto Q(x,B)$ is measurable for any $B \in \Bcal(E')$ and
	\item $B \mapsto Q(x,B)$ is a measure for any $x \in E$.
      \end{enumerate*}
      If $Q(x,E') = 1$ for any $x \in E$, then $Q$ is referred to as a Markov kernel from $E$ to $E'$. In this case, let $\Gamma_Q$ be the following change of probability measure:
      \begin{equation}
	\label{eq:gamma_transform}
	\arraycolsep=1pt\def\arraystretch{1}
	\begin{array}{rl}
	  \Gamma_Q: \Pcal(E) & \to \Pcal(E')
	  \\
	  \mu & \mapsto \Gamma_Q(\mu),
	\end{array}
      \end{equation}
      where $\Gamma_Q(\mu)(\d x') = \mu(Q(\cdot, \d x'))$.
      
      If $Q$ denotes the Markov transition kernel from a space $E_{t-1}$ representing the target state at time $t-1$ to a space $E_t$ corresponding to time $t$, then $\Gamma_Q$ transforms the information maintained on the state of some target from time $t - 1$ to $t$. In a linear Gaussian problem, it describes the usual single-target prediction step of the Kalman filter \cite{Kalman_RE_1960}.
      
    \subsection{Data update} \label{subsec:notation_boltzmann_gibbs_transformation}
      Let $G \in \Lcal^{\infty}(E)$, the following change of measure is referred to as Boltzmann-Gibbs transformation \cite{Del_Moral_P_2009}:
      \begin{equation}
	\label{eq:BoltzmannGibbs}
	\arraycolsep=1pt\def\arraystretch{1}
	\begin{array}{rl}
	  \Psi_G: \Mcal(E) & \to \Pcal(E)
	  \\
	  \eta & \mapsto \Psi_G(\eta),
	\end{array}
      \end{equation}
      where, assuming $\eta(G) > 0$,
      \begin{equation}
	\Psi_G(\eta)(\d x) = \dfrac{1}{\eta(G)} G(x) \eta(\d x). \label{eq:BoltzmannGibbs_detail}
      \end{equation} 
      If $G$ denotes the likelihood function for some observation $z$ produced by a sensor, then $\Psi_G$ transforms the information maintained on the state of some target at time $t$ once it is associated to $z$. In a linear Gaussian problem, it describes the usual single-target/single-observation update step of the Kalman filter.

  \section{Stochastic populations for multi-target estimation problems} \label{sec:stochastic}
    Here we provide the salient features of the general estimation framework for stochastic populations \cite{Houssineau_J_2015_1_2} that we shall exploit for a multi-target estimation problem. For the rest of the paper, we denote by $\Xcal$ a population of individuals of interest, or \emph{targets}, on which some operator wishes to determine individual characteristics gathered into a \emph{target state} (position, velocity, etc.), and we shall assume that the targets behave independently from each other. In this context, one can describe two sources of uncertainties regarding the population of targets: its composition, and the individual state of its members.
    
    \subsection{Multi-target configuration measures} \label{subsec:stochastic_configuration}
      Assume that $E$ denotes a \emph{target state space} suitable for the description of individual target states, and that the operator possesses a set $\{p_i : i \in \Ibb\}$ of probability distributions in $\Pcal(E)$ indexed by some set $\Ibb$, such that $p_i = p_j \Rightarrow i = j$ for any $i, j \in \Ibb$. The \emph{track} $(i, p_i)$ associates a unique \emph{target index} $i \in \Ibb$ to a corresponding description of the target's state in $E$ through the probability distribution (or \emph{law}) $p_i$ of the target. In the context of filtering, as it will be seen with for the \ac{disp} filter in this paper, the composition of the index set $\Ibb$ and the corresponding laws are enriched along the scenario, most notably when new observations are collected from the sensor system observing the scene.
      
      We define a \emph{multiplicity} on $\Ibb$ as a family of integers $\nbs \in \Nbb^{\Ibb}$ indexed by $\Ibb$, describing a composition of the population $\Xcal$ where there are $\nbs_i$ individuals represented by the track $(i, p_i)$, $i \in \Ibb$. It is shown in \cite[chap. 2]{Houssineau_J_2015_1_2} that a suitable representation of the corresponding population is given by the \emph{multi-target configuration measure} $\mu_{\nbs} \in \Mcal(\Pcal(E))$, a integer-valued measure on the space of probability distributions $\Pcal(E)$ defined as
      \begin{equation} \label{eq:configuration_with_integer_measures}
	\mu_{\nbs} = \sum_{i \in \Ibb} \nbs_i \delta[p_i],
      \end{equation}
      where $\delta[p]$ is the Dirac measure at point $p$. The multi-target configuration measure $\mu_{\nbs}$ embodies all the information maintained by the operator on the population $\Xcal$, given the specific composition $\nbs$. The case $\nbs_i \geq 2$ indicates that \emph{several} targets are indexed by the same track $(i, p_i)$: the law $p_i$ describes \emph{collectively} the state of \emph{each} target of the population indexed by $i$, and these targets are \emph{indistinguishable} for the purpose of estimation. Conversely, the case $\nbs_i = 1$ indicates that a \emph{single} target of the population is characterised by the track $(i, p_i)$, and is thus \emph{distinguishable} for the purpose of estimation. The concept of target distinguishability is a key element of the estimation framework for stochastic populations \cite{Houssineau_J_2015_1_2}, and plays an important role in the construction of the \ac{disp} filter.
      
      While the form \eqref{eq:configuration_with_integer_measures} is concise and convenient to describe the propagation of information in the context of filtering, it is shown in \cite[chap. 2]{Houssineau_J_2015_1_2} that $\mu_{\nbs}$ induces an equivalent joint probability measure which enables events regarding the state of individuals to be assessed.
      
      \begin{example}
	Assume $\Ibb = \{\a, \b, \c\}$, the target laws $p_{\a}$, $p_{\b}, p_{\c}$, and the multiplicity $\nbs \in \Nbb^{\Ibb}$ such that $\nbs_{\a} = 1$, $\nbs_{\b} = 2$, $\nbs_{\c} = 0$. The corresponding multi-target configuration $\mu_{\nbs}$ is then
	\begin{equation} \label{eq:configuration_with_joint_probability_measure_ex_1}
	  \mu_{\nbs} = \delta[p_{\a}] + 2\delta[p_{\b}].
	\end{equation}
	Consider some subsets $A, B \in \Bcal(E)$ as regions of the target state space in which one wishes to assess the presence of targets.
	
	Then, according to $\mu_{\nbs}$, the probability that there is one target with index $\a$ lying within $A$, two targets with index $\b$, one lying within $A$ and one within $B$, and no target with index $\c$, is
	\begin{equation} \label{eq:configuration_with_joint_probability_measure_ex_2}
	  p_{\a}(A)p_{\b}(A)p_{\b}(B),
	\end{equation}
	and the probability that there is one target with index $\c$ is equal to zero since the composition of the population does not agree with the multiplicity $\nbs$.
      \end{example}  
      The formal correspondence between a multi-configuration measure $\mu_{\nbs}$ and the induced joint probability measure is provided in Section~\ref{sec:equivalence_mu_P} of the Appendix, though it is not essential for the rest of the paper as the form \eqref{eq:configuration_with_integer_measures} will be exploited from now on.
      
    \subsection{Stochastic population} \label{subsec:stochastic_composition}  
      We have seen in Section~\ref{subsec:stochastic_configuration} that the information on the population for a given multiplicity $\nbs$ is represented by the multi-target configuration measure $\mu_{\nbs}$. The uncertainty on the composition of the population is then embedded in a \emph{random configuration measure} $\Cfrak$, i.e.\ a random variable on the set of integer-valued measures\footnote{In a more general case beyond the scope of this paper \cite{Houssineau_J_2015_1_2}, one might not be able to index the targets through their probability distributions, i.e., the countable index set $\Ibb$ might not be available. In this case, the stochastic population $\Cfrak$ is a general point process on $\Pcal(E)$ which cannot be reduced to a random multiplicity as in Eq.~\eqref{eq:stochastic_population}.} on $\Pcal(E)$, characterised by
      \begin{equation} \label{eq:stochastic_population}
	\Cfrak(F) = \sum_{i \in \Ibb} \Nbs_i F(p_i), 
      \end{equation}
      for any $F \in \Lcal^{\infty}(\Pcal(E))$, where $\Nbs$ is a random variable on $\Nbb^{\Ibb}$ induced by $\Cfrak$. The \emph{stochastic population} $\Cfrak$ thus encapsulates all the uncertainty about the population, as $\Nbs$ describes its composition and the laws $p_i$ describe the states of its members. Denoting by $\cbo$ the probability mass function of $\Nbs$, the law of the stochastic population $\Cfrak$ is given by
      \begin{equation} \label{eq:stochastic_population_law}
	P_{\Cfrak} = \sum_{\nbs \in \Nbb^{\Ibb}} \cbo(\nbs)\delta[\mu_{\nbs}], 
      \end{equation}
      i.e., a realization of the stochastic population $\Cfrak$ is the multi-target configuration $\mu_{\nbs}$, given in Eq.~\eqref{eq:configuration_with_integer_measures}, with probability $\cbo(\nbs)$.
 
      \begin{remark}
	The case where the number of target representations is reduced to one (i.e., $|\Ibb| = 1$) is interesting to consider. The stochastic population \eqref{eq:stochastic_population} then reduces to a random variable $N$ on $\Nbb$, describing the number of targets in the whole population $\Xcal$, and a single law $p \in \Pcal(E)$ describing the state of each of the targets; in other words, the stochastic population $\Cfrak$ becomes equivalent to an \ac{iid} \emph{point process} on $E$ \cite{Stoyan_D_1995, Daley_DJ_2003, Daley_DJ_2008}.
      \end{remark}
      
    \subsection{Population and statistical moments} \label{subsec:stochastic_moment} 
      Similarly to usual random variables, we can produce statistical moments of the stochastic population $\Cfrak$. Since a stochastic population represents two levels of uncertainty regarding the composition of the population and the state of its members, we can define moments of two different nature: the \emph{(full) moments} provide statistics on the composition of the population (e.g. how many targets are represented by some probability distribution $p$), while the \emph{collapsed moments} provide statistics on the state of the targets (e.g. how many targets are lying in some region $B$ of the surveillance scene).\newline    
      
      \subsubsection{Full moments} \label{subsubsec:stochastic_moment_full} 
	The first moment measure ${\M_{\Cfrak} \in \Mcal(\Pcal(E))}$ and the variance $\Var_{\Cfrak} : \Lcal^{\infty}(\Pcal(E)) \to \Rbb$ are defined as
	\begin{align}
	  \M_{\Cfrak}(F) &= \Ebb[\Cfrak(F)],
	  \\
	  \Var_{\Cfrak}(F) &= \Ebb[\Cfrak(F)^2] - \Ebb[\Cfrak(F)]^2,
	\end{align}
	for any $F \in \Lcal^{\infty}(\Pcal(E))$.
	
	The component $\Nbs_i$ in Eq.~\eqref{eq:stochastic_population}, $i \in \Ibb$, is a random variable on $\Nbb$ describing the size of the subpopulation of targets indexed by $i$. Using the law of the population \eqref{eq:stochastic_population_law}, we can write the first moment and covariance of these components as
	\begin{align}
	  \newsubeqblock
	    \mysubeq \mbo_{\Cfrak}(i) &= \Ebb[\Nbs_i]
	    \\
	    \mysubeq &= \sum_{\nbs \in \Nbb^{\Ibb}} \cbo(\nbs) \nbs_i, \label{eq:mean_pop}
	    \\
	  \newsubeqblock
	    \mysubeq \covbo_{\Cfrak}(i, j) &= \Ebb[\Nbs_i\Nbs_j] - \Ebb[\Nbs_i]\Ebb[\Nbs_j]
	    \\
	    \mysubeq &= \sum_{\nbs \in \Nbb^{\Ibb}} \cbo(\nbs)\nbs_i\big[\nbs_j - \mbo_{\Cfrak}(j)\big], \label{eq:var_pop}
	\end{align}
	such that the first moment $\M_{\Cfrak}$ and variance $\Var_{\Cfrak}$ are found to be
	\begin{align}
	  \M_{\Cfrak}(F) &= \sum_{i \in \Ibb} \mbo_{\Cfrak}(i) F(p_i), \label{eq:mean_random_conf}
	  \\
	  \Var_{\Cfrak}(F) &= \sum_{i,j \in \Ibb}\covbo_{\Cfrak}(i, j)F(p_i)F(p_j). \label{eq:var_random_conf}
	\end{align}
	Since the first moment $\M_{\Cfrak}$ and variance $\Var_{\Cfrak}$ are statistical quantities on $\Pcal(E)$, i.e., on the space of probabilities measures on the target state space $E$, they are useful to produce statistics on the \emph{laws} of the targets of the population.
%
	
	\begin{example} \label{ex:moment_close_law}
	  The number of targets represented by a probability distribution within some subset $B \in \Bcal(\Pcal(E))$ has mean and variance given by setting $F = 1_{B}$, i.e.
	  \begin{align} 
	    \M_{\Cfrak}(1_{B}) &= \sum_{i \in \Ibb} \mbo_{\Cfrak}(i) 1_{B}(p_i), \label{eq:mean_random_conf_ex_2}
	    \\
	    \Var_{\Cfrak}(1_{B}) &= \sum_{i,j \in \Ibb}\covbo_{\Cfrak}(i, j)1_{B}(p_i)1_{B}(p_j). \label{eq:var_random_conf_ex_2}
	  \end{align}
	  In particular, considering $B = \{p_i\}$ yields $\mbo_{\Cfrak}(i)$ and $\covbo_{\Cfrak}(i, i)$, i.e., the mean size and variance of the subpopulation of targets indexed by $i \in \Ibb$.
	  
	  \noindent Also, if a metric $d$ is defined on $\Pcal(E)$ then a subset of the form $B_p = \{ p' \in \Pcal(E) \;|\; d(p,p') < \delta \}$, where $\delta \in [0,\infty)$, can be used to estimate the number of targets whose laws are close to $p$ in the sense of $d$.
	\end{example}
      
      \subsubsection{Collapsed moments} \label{subsubsec:stochastic_moment_collapsed} 
	The \emph{collapsed} first moment ${\m_{\Cfrak} \in \Mcal(E)}$ and the variance $\var_{\Cfrak}$ on $\Lcal^{\infty}(E)$ are defined as
	\begin{align}
	  \m_{\Cfrak}(f) &= \M_{\Cfrak}(\chi_f),
	  \\
	  \var_{\Cfrak}(f) &= \Var_{\Cfrak}(\chi_f),
	\end{align}
	where the transformation $\chi_f$ is given by
	\begin{equation}
	  \begin{aligned}
	    \chi_f : \Pcal(E) & \to E
	    \\
	    p & \mapsto p(f),
	  \end{aligned}
	\end{equation}
	for any function $f \in \Lcal^{\infty}(E)$. From Eqs~\eqref{eq:mean_random_conf}, \eqref{eq:var_random_conf}, the collapsed quantities become
	\begin{align} 
	  \m_{\Cfrak}(f) &= \sum_{i \in \Ibb} \mbo_{\Cfrak}(i) p_i(f), \label{eq:mean_random_conf_collapsed}
	  \\
	  \var_{\Cfrak}(f) &= \sum_{i,j \in \Ibb} \covbo_{\Cfrak}(i, j) p_i(f) p_j(f). \label{eq:var_random_conf_collapsed}
	\end{align}
	Since the collapsed first moment $\m_{\Cfrak}$ and variance $\var_{\Cfrak}$ are statistical quantities on the target state space $E$, they are useful to produce statistics on the \emph{states} of the targets of the population.
	
	Assuming that $E$ is the augmented target state space ${\bar{\Xbo} = \{\psi\} \cup \Xbo}$ describing the surveillance scene, where $\psi$ is the state of targets absent from the scene (see Section~\ref{subsec:model_surveillance}):
	\begin{example} \label{ex:moment_close_state}
	  The number of targets lying in some subset ${B \in \Bcal(\bar{\Xbo})}$ has mean and variance given by setting $f = 1_B$, i.e.
	  \begin{align} 
	    \m_{\Cfrak}(1_B) &= \sum_{i \in \Ibb} \mbo_{\Cfrak}(i) p_i(B), \label{eq:mean_random_conf_collapsed_ex_1}
	    \\
	    \var_{\Cfrak}(1_B) &= \sum_{i,j \in \Ibb} \covbo_{\Cfrak}(i, j) p_i(B)p_j(B). \label{eq:var_random_conf_collapsed_ex_1}
	  \end{align}
	  In particular, considering $B = \{\psi\}$ yields the mean and variance of the number of targets absent from the scene.
	\end{example}
	
	\begin{remark}
	  Moments can also be exploited to evaluate events regarding the states of specific subpopulations of $\Cfrak$. For example, the number of targets lying in some region $B \in \Bcal(\bar{\Xbo})$ \emph{and} belonging to a subpopulation indexed by either $i$ or $j$, where $i, j \in \Ibb$, has mean and variance given by $\M_{\Cfrak}(1_{\{p_i, p_j\}} \chi_{1_B})$ and $\Var_{\Cfrak}(1_{\{p_i, p_j\}} \chi_{1_B})$, respectively.
	\end{remark}
	
	Note that the collapsed mean and variance produced in Example~\ref{ex:moment_close_state} are equivalent to the regional statistics for point processes, introduced in \cite{Delande_E_2014_4} in the context of \ac{fisst}-based filters. We will see in Section~\ref{sec:exploitation} that the statistical moments can be exploited in a similar way from the output of the \ac{disp} filter.\newpage
      
    \subsection{Merging of populations and mixture of laws} \label{subsec:stochastic_mixture}
      The number of indexed subpopulations in $\Cfrak$ reflects the complexity of the information maintained by the operator on the population $\Xcal$; in the context of filtering with stochastic populations, the computational cost of the tracking algorithms is likely to increase with the number of subpopulations $|\Ibb|$ through the data association phase (see Section \ref{sec:filter_update}). For this reason, it is sometimes of interest to \emph{merge} two (or more) subpopulations, i.e., replace them by a single subpopulation in $\Cfrak$, whose probabilistic description accounts for those of the subpopulations about to be merged.
      
      Suppose that the operator wishes to merge two subpopulations with respective index $i, j \in \Ibb$, where $i \neq j$. Denote by $i \oplus j$ the index of the merged subpopulation to be constructed. From Eq.~\eqref{eq:stochastic_population} it is straightforward to construct the probability mass function $c_{i \oplus j}$, accounting for the size of the merged subpopulation, as
      \begin{subequations} \label{eq:merged_population_cardinality}
	\begin{align}
	  c_{i \oplus j}(n) &= \Pbb(\Cfrak(1_{\{p_i, p_j\}}) = n)
	  \\
	  &= \Pbb(\Nbs_i + \Nbs_j = n)
	  \\
	  & = \sum_{\substack{\nbs \in \Nbb^{\Ibb} \\ \nbs_i + \nbs_j = n}} \cbo(\nbs),
	\end{align}
      \end{subequations}
      for any $n \in \Nbb$. We now need to determine the \emph{mixture} $p_{i \oplus j} \in \Pcal(E)$ of the probability distributions $p_i, p_j \in \Pcal(E)$, describing the state of the targets in the merged subpopulation and accounting for the laws $p_i$ and $p_j$.
      
      Assume that the merged subpopulation has size $n$, and consider the probability distribution $p^{(n)}_{i \oplus j} \in \Pcal(E)$ given by
      \begin{subequations} \label{eq:merged_population_law_per_cardinality}
	\begin{align}
	  p^{(n)}_{i \oplus j}(f) &= \frac{\Ebb[\Cfrak(1_{\{p_i, p_j\}} \chi_f) \st \Nbs_i + \Nbs_j = n]}{\Ebb[\Cfrak(1_{\{p_i, p_j\}}) \st \Nbs_i + \Nbs_j = n]}
	  \\
	  &\propto \sum_{\substack{\nbs \in \Nbb^{\Ibb} \\ \nbs_i + \nbs_j = n}} \cbo(\nbs)[\nbs_i p_i(f) + \nbs_j p_j(f)], 
	\end{align}
      \end{subequations}
      for any $f \in \Lcal^{\infty}(E)$. The law $p^{(n)}_{i \oplus j}$ averages the laws $p_i, p_j$ about to be mixed; note in particular that if the number of targets originating from one of the subpopulations (say $i$) is preponderant in the possible compositions of the merged subpopulation of size $n$, then the influence of $p_i$ is preponderant in the composition of the mixture $p^{(n)}_{i \oplus j}$.    

      In some situations, it is useful to simplify the structure of the merged subpopulation $i \oplus j$ by approximating it by a simpler \ac{iid} representation collapsing the construction \eqref{eq:merged_population_law_per_cardinality} over the possible sizes, i.e.
      \begin{subequations} \label{eq:merged_population_law_unique}
	\begin{align}
	  p_{i \oplus j}(f) &= \frac{\Ebb[\Cfrak(1_{\{p_i, p_j\}} \chi_f)]}{\Ebb[\Cfrak(1_{\{p_i, p_j\}})]}
	  \\
	  &\propto \mbo_{\Cfrak}(i) p_i(f) + \mbo_{\Cfrak}(j) p_j(f).
	\end{align}
      \end{subequations}
      
      Merging operations are common in the practical implementations of target tracking algorithms, and we will see in Section~\ref{sec:exploitation} that they can be exploited on the \ac{disp} filter in order to simplify its structure towards an approximate version.
    
  \section{Multi-target estimation with the \ac{disp} filter: general concepts and modelling assumptions} \label{sec:model}
    Here we provide a description of a typical surveillance scenario for which the \ac{disp} filter has been developed, through \begin{enumerate*}[label=\roman*)]
      \item the interactions of the targets with the surveillance scene (Section \ref{subsec:model_surveillance}),
      \item the nature of the \textit{a priori} information available to the operator (Section \ref{subsec:model_prior}), and
      \item the interactions of the sensor system with the surveillance scene (Section \ref{subsec:model_sensor}).
    \end{enumerate*}
    
    \subsection{Surveillance activity} \label{subsec:model_surveillance}
      In the context of this paper, an operator wishes to gather information about some population $\Xcal$ of targets (e.g. vehicles) while they lie in some defined region of the physical space called the \emph{surveillance scene} (e.g. the surroundings of a facility) and over a given period of time. While the existence of each individual in $\Xcal$ is assumed certain throughout the scenario, its presence in the surveillance scene \emph{is not}: any target may possibly enter/leave the surveillance scene during the scenario, such that its presence or absence in the scene at any time is unknown to the operator and part of the estimation problem.
      
      The time flow is indexed by some integer $t$. At any time ${t \geq 0}$, each target within the surveillance scene is represented via a \emph{state} $x$ belonging to some \emph{target state space} $\Xbo_t$, describing physical and measurable characteristics which are unknown and of interest to the operator. The nature of the target space space depends on the nature of the surveillance activity, but is assumed to be a bounded subset of $\Rbb^{d_t}$, such that any state $x \in \Xbo_t$ is a real vector of (finite) dimension $d_t$, with a mixture of continuous components such as position and/or velocity coordinates, and discrete components such as a vehicle type, a level of threat, etc. Conversely, the individuals currently absent from the surveillance scene assume the ``empty state'' $\psi$; in consequence, each target in $\Xcal$ is represented by some point in the \emph{augmented target state space} $\bar{\Xbo}_t = \{\psi\} \cup \Xbo_t$.\footnote{The nature of the state space may evolve across time if the operator wishes, for example, to estimate additional characteristics or modify the physical boundaries of the surveillance scene at some point during the scenario.}

      The goal of the surveillance activity is then for the operator to estimate, at any time $t \geq 0$, the state of \emph{every} target of the population $\Xcal$ in the augmented state space, i.e.:
      \begin{enumerate*}[label=\roman*)]
	\item whether the target is currently in the scene, and
	\item if so, its state.
      \end{enumerate*}
      We shall consider the following assumptions:
      \begin{assumption}{Targets in the population $\Xcal$:}
	\begin{enumerate}[labelindent = \parindent, leftmargin = *, series = assumption, label = (M\arabic*)]
	  \item behave independently; \label{mod:individual_independence}
	  \item enter the scene at most once during the scenario; \label{mod:individual_goulp}
	  \item all have state $\psi$ before $t = 0$; \label{mod:individual_preexisting}
	  \item their evolution is Markovian between two successive time steps.\label{mod:transition_individual}
	\end{enumerate}
      \end{assumption}  
      Assumption \ref{mod:individual_goulp} implies that if the same target enters twice in the surveillance scene during the scenario, it will be treated as two different objects for estimation purposes. Following these assumptions, we can divide the population at any time $t \geq 0$ as
      \begin{equation}
	\Xcal_t = \Xcal^{\psibo}_t \cup \Xcal^{\a}_t \cup \Xcal^{\p}_t, \label{eq:population_decomposition}
      \end{equation}
      where
      \begin{enumerate*}[label=\roman*)]
	\item $\Xcal^{\psibo}_t$ are the targets who have not entered the scene yet (and may never do so),
	\item $\Xcal^{\a}_t$ are the \emph{appearing} targets, i.e., those who have just entered the scene,
	\item $\Xcal^{\p}_t$ are the \emph{persistent} targets, i.e., those who have entered the scene in a previous time step (and may have already left it).
      \end{enumerate*}
      
    \subsection{Prior information} \label{subsec:model_prior}
      At any time $t \geq 0$, the operator may possess and exploit some information on the population $\Xcal^{\a}_t$ of appearing individuals (e.g. individuals are likely to appear in groups, alongside roads or from the North, etc.). We shall assume that:
      \begin{assumption}{Prior information (time $t \geq 0$):}
	\begin{enumerate}[labelindent = \parindent, leftmargin = *, resume = assumption, label=(M\arabic*)]
	  \item the prior information on the appearing targets $\Xcal^{\a}_t$ can be described by a single population of \emph{indistinguishable} individuals, independent from the persistent targets $\Xcal^{\p}_t$. \label{mod:information_appearing}
	\end{enumerate}
      \end{assumption}
      
      Assumption~\ref{mod:information_appearing} states that the targets appearing at a common time are indistinguishable; they will remain so until detected by the sensor system. Lifting this assumption would allow for a more refined description of appearing targets, and lead to a more general filtering solution. 
      
    \subsection{Sensor system and observation process} \label{subsec:model_sensor}
      At any time $t \geq 0$, the surveillance scene is observed by some sensor system, providing information on the targets through a collection of \emph{observations} (or \emph{measurements}). An observation  $z$ belongs to some observation space $\Zbo_t$ and describes physical quantities measurable by the sensor system and relevant to a target's individual characteristics of interest (e.g. range, bearing, angle velocity, etc.). The observation space is assumed to be some bounded subset of $\Rbb^{d'_t}$, such that any observation $z \in \Zbo_t$ is a real vector of (finite) dimension $d'_t$. The observations may have a mixture of continuous and discrete components in the most general form, although a discrete observation space is sufficient to interpret the output of some sensors (e.g. resolution cells for a radar).\footnote{The observation space may not be constant throughout the scenario depending on the context, e.g. if sensors of different nature are exploited at different times.} The measurements collected by the operator may also contain spurious observations or \emph{false alarms}. We shall assume that:
        
      \begin{assumption}{Observation process ($t \geq 0$):}
	\begin{enumerate}[labelindent = \parindent, leftmargin = *, resume = assumption, label=(M\arabic*)]
	  \item a target produces at most one observation (if not, it is a \emph{missed detection}); \label{mod:observation_per_target}
	  \item an observation originates from at most one target (if not, it is a \emph{false alarm}); \label{mod:target_per_observation}
	  \item targets outside the scene produce no observations; \label{mod:observation_goulp}
	  \item observations are produced independently; \label{mod:observation_independence}
	  \item observations are distinct, and their number is finite; \label{mod:observation_set}
	  \item an observation produced from a target depends only on its current state.\label{mod:observation_process}
	\end{enumerate}
      \end{assumption}
      Assumptions \ref{mod:observation_per_target} and \ref{mod:target_per_observation} are a central element of the \ac{disp} filter: they state that individual observations provide specific information about individual targets, and that targets become distinguishable through sensor detections (see discussion on target distinguishability in Section~\ref{subsec:stochastic_configuration}). Lifting Assumption~\ref{mod:target_per_observation} means that targets never become distinguishable and leads to the construction of a more general filtering solution \cite[chap. 3]{Houssineau_J_2015_1_2}.       
      
      Assumption~\ref{mod:observation_set} guarantees that the observations collected by the operator at any time $t \geq 0$ can be described by a (possibly empty) \emph{observation set} $Z_t$. In order to account for the missed detections, it is convenient to introduce the ``empty observation'' $\phi$ and define, at any time $t \geq 0$,
      \begin{enumerate*}[label=\roman*)]
	\item the \emph{augmented observation set} $\bar{Z}_t = \{\phi\} \cup Z_t$, and
	\item the \emph{augmented observation space} $\bar{\Zbo}_t = \{\phi\} \cup \Zbo_t$.
      \end{enumerate*} 
      Assumption~\ref{mod:observation_goulp} implies that no information is acquired at time $t \geq 0$ on the population $\Xcal^{\psibo}_t$ of targets that have not entered the scene yet. In consequence, the \ac{disp} filter aims at estimating the remaining population $\Xcal_t \setminus \Xcal^{\psibo}_t$, i.e., the targets that have already entered the scene. As an alternative to Eq.~\eqref{eq:population_decomposition}, we can divide the population at any time $t \geq 0$ as
      \begin{equation}
	\Xcal_t = \Xcal^{\psibo}_t \cup \Xcal^{\circ}_t \cup \Xcal^{\bullet}_t, \label{eq:population_decomposition_further}
      \end{equation}
      where
      \begin{enumerate*}[label=\roman*)]
	\item $\Xcal^{\circ}_t$ are the targets that have already entered the scene, but have never been detected so far (also called \emph{yet-to-be-detected} targets),
	\item $\Xcal^{\bullet}_t$ are the targets that have been detected at least once (also called \emph{previously-detected} targets).
      \end{enumerate*}

  \section{DISP filter: target representation} \label{sec:filter_representation}
    In this section, we highlight the articulation between the structure of the \ac{disp} filter and the concepts of multi-target configuration measures and stochastic populations presented in Section~\ref{sec:stochastic}. In this section, $t \geq 0$ represents an arbitrary time step relevant to the scenario. As established in Section~\ref{sec:model}, the operator aims to estimate the targets that have entered the scene so far, relying on prior information on the appearing targets $\Xcal^{\a}_{t'}$ and the collected observations $Z_{t'}$ during all the past times $0 \leq t' \leq t$.
    
    \subsection{Observation paths} \label{subsec:filter_representation_observation_path}
\begin{figure}[H]
	\centering
	\includegraphics{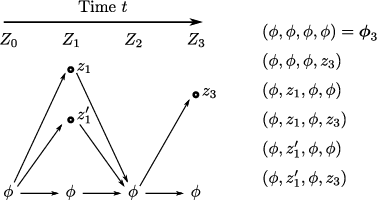}
	\caption{The observation paths at time $t = 3$, given a sequence of collected measurements. The six possible observation paths are listed on the right.\label{fig:observation_path}}	
      \end{figure}

      We first introduce the set $\bar{\Ybo}_t$ of observation histories\ManuComment{reference to DelMoral's and Pace's work here?}, or \emph{observation paths} up to time $t$, defined as
      \begin{equation} \label{eq:observation_path_set}
	\bar{\Ybo}_t = \bar{Z}_0 \times \ldots \times \bar{Z}_t,
      \end{equation}
      as well as the set $\Ybo_t$ of non-empty observation paths defined as $\Ybo_t = \bar{\Ybo}_t \setminus \{\phibs_t\}$, where $\phibs_t$ is the \emph{empty observation path} at time $t$, i.e.\ the sequence made of $t+1$ empty observations $\phi$. Some examples of observation paths are illustrated in Figure~\ref{fig:observation_path}. 

    \subsection{Previously-detected targets} \label{subsec:filter_representation_distinguishable}
      As we have seen in Section~\ref{subsec:model_sensor} each observation (if not a false alarm) provides specific information on one target, and thus the previously-detected targets $\Xcal^{\bullet}_t$ are distinguishable for the purpose of estimation. A target in $\Xcal^{\bullet}_t$ can then be indexed by the pair $i = (t_{\a},y)$, made of its observation path $y \in \Ybo_t$ as well as its time of appearance in the scene $0 \leq t_{\a} \leq t$ (also called ``time of arrival'') and can be described by some associated probability distribution $p^i_t \in \Pcal(\bar{\Xbo}_t)$, describing the current state of the target indexed by $i$ and \emph{no one else}.
      
      Note that the probability distribution $p^i_t$ is defined on the augmented state space $\bar{\Xbo}_t$. The scalar $p^i_t(\Xbo_t)$ denotes the probability that the target indexed by $i$ is currently in the scene, given that it exists; it is called the \emph{probability of presence} of target $i$. The restriction of the probability distribution $p^i_t$ to the state space $\Xbo_t$ is called the \emph{spatial distribution} of the target $i$.

      \JeremComment{Remark somewhere about the fact that a survival AND a death kernel can be defined separately, but that increases substantially the number of hypotheses. However, mixing these two kernels also prevent from estimating the time of death.} 
      The set of all possible indices for previously-detected targets is given by
      \begin{equation} \label{eq:index_distinguishable}
	\Ibb^{\bullet}_t = \{(t', y) \st y \in \Ybo_t, 0 \leq t' \leq t\}.
      \end{equation}

      It is clear from the construction of the observation paths \eqref{eq:observation_path_set} that each target in $\Ibb^{\bullet}_t$ is a valid characterization of \emph{one} individual in $\Xcal^{\bullet}_t$. However, not every subset of $\Ibb^{\bullet}_t$ forms a valid characterization of \emph{the} individuals $\Xcal^{\bullet}_t$, since any two individuals whose observation paths are \emph{incompatible}, i.e., who share a non-empty observation\footnote{In the situation illustrated in Figure~\ref{fig:observation_path}, for example, $(\phi, \phi, \phi, z_3)$ and $(\phi, z_1, \phi, \phi)$ are compatible, but $(\phi, \phi, \phi, z_3)$ and $(\phi, z_1, \phi, z_3)$ are not since they share measurement $z_3$.}, may not exist simultaneously without violating Assumption~\ref{mod:target_per_observation}. The subsets of targets in $\Ibb^{\bullet}_t$ made of pairwise-compatible targets are called \emph{hypotheses} and form a set $\Hbo_t$ maintained by the \ac{disp} filter\footnote{The proof is given in Section~\ref{subsec:filter_update_compatibility}.}.
      
      Following Eq.~\eqref{eq:configuration_with_integer_measures} and given some hypothesis $H \in \Hbo_t$, the case where the previously-detected targets $\Xcal^{\bullet}_t$ are characterised by the indices in $H$ is then described by the multi-target configuration measure
      \begin{equation}
	\mu^H_t = \sum_{i \in H} \delta[p^i_t]. \label{eq:configuration_distinguishable}
      \end{equation}

    \subsection{Yet-to-be-detected targets} \label{subsec:filter_representation_indistinguishable}
      As we have seen in Section~\ref{subsec:model_prior} the targets appearing in the scene simultaneously are indistinguishable for the purpose of estimation, and remain so until their (eventual) first detection. The subpopulation of targets that appeared at time $0 \leq t' \leq t$ and have never been detected so far, undistinguishable from each other, are represented with a \emph{single} index $(t',\phibs_t)$, associated to some probability distribution $p^{(t',\phibs_t)}_t \in \Pcal(\bar{\Xbo}_t)$. In other words, a \emph{single} track describes \emph{collectively} the state of \emph{each} target in this subpopulation. The notions of probability of presence and spatial distribution defined in Section~\ref{subsec:filter_representation_distinguishable} hold for indistinguishable targets as well.
      
      Note that yet-to-be detected targets appearing at different time steps $t', t''$ are characterised by distinct indices, and described by the distinct probability distributions $p^{(t', \phibs_t)}_t$ and $p^{(t'', \phibs_t)}_t$, respectively. The set of all possible indices for yet-to-be-detected targets is given by
      \begin{equation} \label{eq:index_indistinguishable}
	\Ibb^{\circ}_t = \{(t', \phibs_t) \st 0 \leq t' \leq t\}.
      \end{equation}

      Following Eq.~\eqref{eq:configuration_with_integer_measures} and given a family of integers ${\nbs \in \Nbo_t = \Nbb^{[0,t]}}$ indexed by times between $0$ and $t$, the case where the yet-to-be detected targets $\Xcal^{\circ}_t$ are composed of $\nbs_{t'}$ individuals that appeared at time $t'$, $0 \leq t' \leq t$, is then described by the multi-target configuration measure
      \begin{equation} \label{eq:configuration_indistinguishable}
	\mu^{\nbs}_t = \sum_{0 \leq t' \leq t} \nbs_{t'} \delta\Big[ p^{(t',\phibs_t)}_t \Big].
      \end{equation}
      
      \begin{remark}
	The special case $\nbs_{t'} = 1$ in Eq.~\eqref{eq:configuration_indistinguishable} represents the situation where the subpopulation of yet-to-be-detected targets that appeared at some time $0 \leq t' \leq t$ is reduced to a single individual, which therefore becomes distinguishable (see Section~\ref{subsec:stochastic_configuration}). This has no consequence on the construction of the \ac{disp} filter and, for the sake of simplicity, we shall refer to the yet-to-be-detected targets as indistinguishable in the rest of the paper. 
      \end{remark}
      
    \subsection{Law of the population} \label{subsec:filter_representation_population}
      Following Sections~\ref{subsec:filter_representation_distinguishable} and \ref{subsec:filter_representation_indistinguishable}, the set of all possible target indices is given by
      \begin{subequations}
	\begin{align}
	  \Ibb_t &= \Ibb^{\circ}_t \cup \Ibb^{\bullet}_t
	  \\
	  &= \{(t', y) \st y \in \bar{\Ybo}_t, 0 \leq t' \leq t\}.
	\end{align}
      \end{subequations}
      
\begin{figure}[ht]
	\centering
	\includegraphics{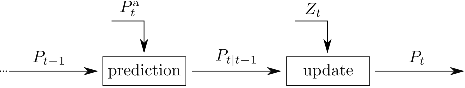}
	\caption{Data flow of the \ac{disp} filter (time $t > 0$). $P^{\a}_{t}$ is a model parameter describing the appearing targets (detailed in Section~\ref{sec:filter_prediction}) and the observation set $Z_t$ is collected from the sensor system.\label{fig:data_flow}} 
      \end{figure}

      The probability mass function on $\Hbo_t \times \Nbo_t$ describing the composition of the population $\Xcal$ is denoted $\wbo_t$: for instance, $\wbo_t(H,\nbs)$ is the probability that the population $\Xcal$ is represented by the \emph{configuration} $(H, \nbs)$, i.e., the probability that the previously-detected targets are those indexed in $H \in \Hbo_t$, together with $\nbs_{t'} \in \Nbb$ yet-to-be-detected targets that appeared at time $t'$, $0 \leq t' \leq t$. Following Eq.~\eqref{eq:stochastic_population_law} the stochastic population $\Cfrak_t$ maintained by the \ac{disp} filter is thus described by the law
      \begin{equation} \label{eq:population_law}
	P_t = \sum_{H \in \Hbo_t} \sum_{\nbs \in \Nbo_t} \wbo_t(H,\nbs) \delta\big[ \mu^H_t + \mu^{\nbs}_t \big],
      \end{equation}
      and the data flow of the \ac{disp} filter is as depicted in Figure~\ref{fig:data_flow}.
      
  \section{DISP filter: time prediction} \label{sec:filter_prediction}        
    In this section we provide a detailed construction of the Bayes prediction step of the \ac{disp} filter, i.e., we describe how the law of the population is updated from the posterior $P_{t-1}$ to the prediction $P_{t|t-1}$ (see Figure~\ref{fig:data_flow}). In this section, $t \geq 0$ designs an arbitrary time step relevant to the scenario. The proof of the \ac{disp} prediction can be found in \cite[chap. 3]{Houssineau_J_2015_1_2}.
    
    \subsection{Input} \label{subsec:filter_prediction_input}  
      The current input is the law $P_{t-1}$ of the form \eqref{eq:population_law}, representing the targets who have entered the scene no later than at time $t-1$. The previously-detected targets are indexed with the set $\Ibb^{\bullet}_{t-1}$ of the form \eqref{eq:index_distinguishable}, while the yet-to-be-detected ones are indexed with the set $\Ibb^{\circ}_{t-1}$ of the form \eqref{eq:index_indistinguishable}.
      
      The case $t = 0$ deserves special mention. Given Assumption~\ref{mod:individual_preexisting} none of the targets have entered the scene so far, the set of hypotheses $\Hbo_{-1}$ is thus reduced to the singleton
      \begin{equation} \label{eq:initialisation_hypothesis}
	\Hbo_{-1} = \{\emptyset\}.
      \end{equation}
      
    \subsection{Modelling} \label{subsec:filter_prediction_model}
      Given Assumption~\ref{mod:transition_individual} the knowledge of the operator regarding the evolution of the targets since time $t-1$ is given by a Markov kernel $m_{t-1,t}$ from the previous target state space $\bar{\Xbo}_{t-1}$ to the current one $\bar{\Xbo}_t$. In particular, the function $m_{t-1,t}(\cdot, \{\psi\})$ describes the transition for a target to the empty state $\psi$; since Assumption~\ref{mod:individual_goulp} implies that no target may re-enter the surveillance scene, it holds that
      \begin{equation}
	m_{t-1, t}(\psi, \{\psi\}) = 1. \label{eq:prediction_goulp}
      \end{equation}
      In target tracking applications it is then customary to rewrite $m_{t-1,t}$ through a reduced Markov kernel $\hat{m}_{t-1,t}$ from $\Xbo_{t-1}$ to $\Xbo_t$ and a \emph{probability of survival} $p_{\s, t}$ on $\Xbo_{t-1}$ such that
      \begin{equation} \label{eq:prediction_transition}
	\begin{dcases}
	  m_{t-1,t}(x, \d x') = p_{\s, t}(x)\hat{m}_{t-1,t}(x, \d x'), \!\!\!\!& x \in \Xbo_{t-1}, x' \in \Xbo_t,
	  \\
	  m_{t-1,t}(x, \{\psi\}) = 1 - p_{\s, t}(x), & x \in \Xbo_{t-1},
	  \\
	  m_{t-1,t}(\psi, \d x') = 0, & x' \in \Xbo_t,
	  \\
	  m_{t-1,t}(\psi, \{\psi\}) = 1.
	\end{dcases}
      \end{equation}
      The reduced Markov kernel $\hat{m}_{t-1,t}$ describes the transition for a target within the surveillance scene. In most practical problems it accounts for the operator's knowledge about the motion of targets in the physical space -- presence of roads and/or obstacles, maximum speed per vehicle type, etc. The probability of survival $p_{\s, t}(x)$ is a scalar that describes how likely is a target, with state $x$ at time $t - 1$, to be still in the scene at time $t$.
      
      Given Assumption~\ref{mod:information_appearing}, the prior information on the appearing targets $\Xcal^{\a}_t$ is represented by a subpopulation of indistinguishable targets indexed with $i_{\a} = (t, \phibs_{t-1})$. Their state is described by some probability distribution $p^{i_{\a}}_{t|t-1} \in \Pcal(\bar{\Xbo}_t)$, and their number by some probability mass function $\rho^{\a}_t$ on $\Nbb$, both parameters of the \ac{disp} filter (see Figure~\ref{fig:data_flow}). Note that, by construction, appearing targets are currently in the surveillance scene. It thus holds that
      \begin{equation} \label{eq:probability_presence_appearing_target}
	p^{i_{\a}}_{t|t-1}(\Xbo_t) = 1,
      \end{equation}
      that is, an appearing target is present in the scene \textit{almost surely}.
      
      The case where there are $n \geq 0$ appearing individuals is thus represented by the multi-target configuration measure
      \begin{equation}
	\mu^{i_{\a},n}_{t|t-1} = n \delta[p^{i_{\a}}_{t|t-1}], \label{eq:configuration_appearing}
      \end{equation}
      and following Eq.~\eqref{eq:stochastic_population_law} the population of appearing individuals is then described by the stochastic population $\Cfrak^{\a}_t$ whose law is given by
      \begin{equation} \label{eq:appearing_law}
	P^{\a}_t = \sum_{n \geq 0} \rho^{\a}_t(n)\delta\Big[\mu^{i_{\a},n}_{t|t-1}\Big].
      \end{equation} 
      
    \subsection{Target prediction} \label{subsec:filter_prediction_track}
      The information gathered so far on any target $i \in \Ibb_{t-1}$ is described by a probability distribution $p^i_{t-1}$ on the former state space $\bar{\Xbo}_{t-1}$; in the prediction step, it is transferred to the current state space $\bar{\Xbo}_t$ through the Markov kernel $m_{t-1,t}$~\eqref{eq:prediction_transition}. The resulting probability distribution $p^i_{t|t-1} \in \Pcal(\bar{\Xbo}_t)$ is found to be \cite{Houssineau_J_2015_2}
      \begin{equation}
	p^i_{t|t-1} = \Gamma_{m_{t-1,t}}(p^i_{t-1}), \label{eq:prediction_track}
      \end{equation}
      that is, as detailed in Section~\ref{subsec:notation_markov_kernel},
      \begin{equation}
	p^i_{t|t-1}(\d x) = \int_{\bar{\Xbo}_{t-1}}\!\!\!\!\!m_{t-1,t}(x', \d x) p^i_{t-1}(\d x'). \label{eq:prediction_track_detail}
      \end{equation}
      Note that the prediction step \eqref{eq:prediction_track_detail} applies to all targets, regardless of their distinguishability. It is formative to write explicitly the evolution of the probability of presence of a target in the scene:
      \begin{property} \label{prop:prediction_probability_presence}
	Let $i \in \Ibb_{t-1}$. The probability of presence of the predicted target $i$ is found to be
	\begin{equation} \label{eq:predicted_track_probability_presence}
	  p^i_{t|t-1}(\Xbo_t) = p^i_{t-1}(\Xbo_{t-1}) - \int_{\Xbo_{t-1}}\!\!\!\!\![1 - p_{\s, t}(x)]p^i_{t-1}(\d x).
	\end{equation}
      \end{property}
      The proof is given in Appendix~\ref{subsec:proof_prop_prediction_probability_presence}. Given Assumption~\ref{mod:individual_goulp} no target can re-enter the scene and thus, as confirmed by Eq.~\eqref{eq:predicted_track_probability_presence}, the probability of presence of a target $i$ is non-increasing during the prediction step. It is monotonous \ac{iff} the probability of survival is one over the support of $p^i_{t-1}$ -- that is, \ac{iff} the target $i$ stays in the scene \textit{almost surely} -- and it drops to zero \ac{iff} the probability of survival is zero over the support of $p^i_{t-1}$ -- that is, \ac{iff} the target $i$ leaves the scene \textit{almost surely}.
      
    \subsection{Population prediction} \label{subsec:filter_prediction_hypothesis}
      The appearing targets are, by construction, yet-to-be-detected; the index set $\Ibb^{\circ}_{t-1}$ is thus augmented as follows
      \begin{equation} \label{eq:population_ind_prediction}
	\Ibb^{\circ}_{t|t-1} = \Ibb^{\circ}_{t-1} \cup \{(t, \phibs_{t-1})\}.
      \end{equation}
      On the other hand, since the observation set $Z_t$ is not available yet (see Figure~\ref{fig:data_flow}), neither the distinguishable targets indexed in $\Ibb^{\bullet}_{t-1}$ nor the composition of the hypotheses in $\Hbo_{t-1}$ are modified by the prediction step. The predicted set of target indices is thus
      \begin{equation} \label{eq:prediction_index_total}
	\Ibb_{t|t-1} = \Ibb^{\bullet}_{t-1} \cup \Ibb^{\circ}_{t|t-1}.
      \end{equation}
    
      Since the subpopulations of appearing and persistent targets are assumed independent, the probability mass function on $\Hbo_{t-1} \times \Nbo_t$, for any hypothesis $H \in \Hbo_{t-1}$ and any family of integers $\hat{\nbs} \in \Nbo_t$, becomes
      \begin{equation} \label{eq:prediction_mass_function}
	\wbo_{t|t-1}(H,\hat{\nbs}) = 
	\begin{dcases}
	  \wbo_{t-1}(H,\nbs) \rho^{\a}_t(\hat{\nbs}_t), &t > 0
	  \\
	  \rho^{\a}_0(\hat{\nbs}_0), &t = 0,
	\end{dcases}
      \end{equation}
      where $\nbs \in \Nbo_{t-1}$ is the family of integers such that $\nbs_{t'} = \hat{\nbs}_{t'}$ for all $0 \leq t' < t$.

    \subsection{Output} \label{subsec:filter_prediction_output}
      Following Eq.~\eqref{eq:prediction_track} an hypothesis $H \in \Hbo_{t-1}$ is now described by the multi-target configuration measure
      \begin{equation} \label{eq:prediction_hypothesis}
	\mu^H_{t|t-1} = \sum_{i \in H} \delta[p^i_{t|t-1}],
      \end{equation}
      while the subpopulation of yet-to-be detected targets composed of $\nbs_{t'}$ individuals that appeared at time $t'$, $0 \leq t' \leq t$, is now described by the multi-target configuration measure
      \begin{equation} \label{eq:prediction_undetected}
	\mu^{\nbs}_{t|t-1} = \sum_{0 \leq t' \leq t} \nbs_{t'} \delta\Big[ p^{(t',\phibs_{t-1})}_{t|t-1} \Big].
      \end{equation}
      The law of the stochastic population $\Cfrak_{t|t-1}$ maintained by the \ac{disp} filter thus becomes
      \begin{equation} \label{eq:prediction_total}
	P_{t|t-1} = \sum_{H \in \Hbo_{t-1}} \sum_{\nbs \in \Nbo_t} \wbo_{t|t-1}(H,\nbs) \delta\big[\mu^H_{t|t-1} + \mu^{\nbs}_{t|t-1}\big].
      \end{equation}
	
  \section{DISP filter: data update} \label{sec:filter_update}
    In this section we provide a detailed construction of the Bayes update step of the \ac{disp} filter, i.e., we describe how the law of the population is updated from the prediction $P_{t|t-1}$ to the posterior $P_t$ (see Figure~\ref{fig:data_flow}). In this section, $t \geq 0$ designs an arbitrary time step relevant to the scenario, and $Z_t$ an arbitrary observation set collected from the sensor system. The proof of the \ac{disp} update can be found in \cite[chap. 3]{Houssineau_J_2015_1_2}.
      
    \subsection{Modelling} \label{subsec:filter_update_model}
      Given Assumption~\ref{mod:observation_process} the knowledge of the operator about the observation process, and in particular the imperfections of the sensor (measurement noise, false alarms, missed detections) is described by
      \begin{enumerate*}[label=\roman*)]
	\item a non-negative real-valued function $g_t(\cdot, z) \in \Lcal^{\infty}(\bar{\Xbo}_t)$, interpreted as a likelihood and given for any $z \in \bar{\Zbo}_t$, and
	\item a \emph{probability of false alarm} $p_{\fa, t}$ on $\Zbo_t$.
      \end{enumerate*}
      In particular, the function $g_t(\psi, \cdot)$ describes the observation from an target absent from the scene; since Assumption~\ref{mod:observation_goulp} implies that no target can be detected unless it is in the scene, it holds that
      \begin{equation} \label{eq:update_goulp}
	\begin{dcases}
	  g_t(\psi, z) = 0, & z \in \Zbo_t,
	  \\
	  g_t(\psi, \phi) = 1.
	\end{dcases}
      \end{equation}
      For any observation $z \in \Zbo_t$, it is then customary to write $g_t(\cdot, z)$ through a restricted likelihood $\ell_t(\cdot, z) \in \Lcal^{\infty}(\Xbo_t)$ and a \emph{probability of detection} $p_{\d, t}$ on $\bar{\Xbo}_t$ such that
      \begin{equation} \label{eq:update_likelihood}
	\begin{dcases}
	  g_t(x, z) = p_{\d, t}(x)\ell_t(x, z), &x \in \Xbo_t, z \in \Zbo_t,
	  \\
	  g_t(x, \phi) = 1 - p_{\d, t}(x), &x \in \Xbo_t,
	  \\
	  g_t(\psi, z) = 0, &z \in \Zbo_t,
	  \\
	  g_t(\psi, \phi) = 1.
	\end{dcases}
      \end{equation}
      The restricted likelihood describes the accuracy of the sensor -- $\ell_t(x, z)$ denotes the likelihood that an observation $z$ originates from a target with state $x$. The probability of detection $p_{\d, t}(x)$ is a scalar that describes how likely a target with state $x$ is to be detected by the sensor\footnote{The support of the function $p_{\d, t}$, i.e., the subset of the target state space $\Xbo_t$ where the probability of detection is non-zero, is called the (current) \emph{field of view} of the sensor. Note that the field of view is included in, but does not necessarily equal to, the surveillance scene: if the sensor coverage is limited, there are ``blind zones'' in which targets cannot be observed.} at the current time $t$.
      
      In the context of this paper, the false alarms are spurious observations stemming from the sensor system itself, independently of the population $\Xcal$ (e.g. malfunctions from the sensor device, transmission errors, etc.). For any collected observation $z \in Z_t$, the scalar $p_{\fa, t}(z)$ is the probability that $z$ is a spurious observation; conversely, the scalar $1 - p_{\fa, t}(z)$ is the probability that $z$ originates from a target in $\Xcal$.
      
    \subsection{Target update} \label{subsec:filter_update_target}
      Since a new observation set $Z_t$ is available, the targets' observation path can be updated with a new \emph{data association}, matching a predicted target to an observation in $\bar{Z}_t$, and their probability distribution can be updated accordingly. By construction, the yet-to-be-detected targets are now those who remain to be detected following the current observation, i.e.\footnote{``$\concat$'' is the concatenation operator on sequences, i.e., $(e_1,\dots,e_n)\concat e = (e_1,\dots,e_n,e)$.}
      \begin{subequations} \label{eq:update_index_indistinguishable}
	\begin{align}
	  \Ibb^{\circ}_t &= \{(t', \phibs_{t-1}\concat \phi) \st (t', \phibs_{t-1}) \in \Ibb^{\circ}_{t|t-1}\}
	  \\
	  &= \{(t', \phibs_t) \st (t', \phibs_{t-1}) \in \Ibb^{\circ}_{t|t-1}\}.
	\end{align}
      \end{subequations}
      Conversely, the previously-detected targets are now those who have been at least detected once following the current observation, whether they had already been detected in the past (and hence were already distinguishable), or have just been detected for the first time (and hence have just become distinguishable), i.e.
      \begin{equation} \label{eq:update_index_distinguishable}
        \Ibb^{\bullet}_t = \{(t', y\concat z) \st (t', y) \in \Ibb_{t|t-1}, z \in \bar{Z}_t, y\concat z \neq \phibs_t\}.
      \end{equation}
      The updated set of target indices is thus
      \begin{subequations} \label{eq:update_index_total}
	\begin{align}
	  \Ibb_t &= \Ibb^{\bullet}_t \cup \Ibb^{\circ}_t
	  \\
	  &= \{(t', y\concat z) \st (t', y) \in \Ibb_{t|t-1}, z \in \bar{Z}_t\}.
	\end{align}
      \end{subequations}
      For any target $i = (t', y) \in \Ibb_{t|t-1}$, whether it is associated to an observation ($z \in Z_t$) or a missed detection ($z = \phi$), we shall exploit the slight abuse of notation $i\concat z$ to denote the updated target $(t', y\concat z) \in \Ibb_t$. The probability distribution $p^i_{t|t-1}$ updates following Bayes' rule through the Boltzmann-Gibbs transformation \cite{Del_Moral_P_2009}
      \begin{equation}
	p^{i\concat z}_t = \Psi_{g_t(\cdot, z)}(p^i_{t|t-1}), \label{eq:update_target}
      \end{equation}
      that is, as detailed in Section~\ref{subsec:notation_boltzmann_gibbs_transformation},
      \begin{equation}
	p^{i\concat z}_t(\d x) = \frac{g_t(x, z)p^i_{t|t-1}(\d x)}{\int_{\bar{\Xbo}_t}g_t(x', z)p^i_{t|t-1}(\d x')}. \label{eq:update_target_detail}
      \end{equation}
      It is formative to write explicitly the evolution of the probability of presence of an updated target:
      \begin{property} \label{prop:update_probability_presence}
	Let $i = (t',y) \in \Ibb_{t|t-1}$ and $z \in \bar{Z}_t$. The probability of presence of the updated target $i\concat z = (t', y\concat z) \in \Ibb_t$ is found to be
	\begin{equation} \label{eq:updated_target_probability_presence}
	  p^{i\concat z}_t(\Xbo_t) =
	  \begin{dcases}
	    1, & z \in Z_t,
	    \\
	    \frac{p^i_{t|t-1}(g_t(\cdot, \phi)1_{\Xbo_t})}{1 - p^i_{t|t-1}(\Xbo_t) + p^i_{t|t-1}(g_t(\cdot, \phi)1_{\Xbo_t})}, & z = \phi.
	  \end{dcases}
	\end{equation}
      \end{property}
      The proof is given in Appendix~\ref{subsec:proof_prop_update_probability_presence}. Given Assumption~\ref{mod:observation_goulp} only targets in the scene can be detected and thus, as confirmed by Eq.~\eqref{eq:updated_target_probability_presence}, a target detected this time step lies in the scene \textit{almost surely}. On the other hand, there is no fresh evidence on the presence of targets that are not detected this time step, and their probability of presence is non-increasing.
      
    \subsection{Population update} \label{subsec:filter_update_population}
      Following the construction of the updated targets the composition of the population is reassessed, i.e., 
      \begin{enumerate*}[label=\roman*)]
	\item the set of hypotheses $\Hbo_{t-1}$ is updated to reflect the combinations of compatible previously-detected targets,
	\item the probability mass function $\wbo_{t|t-1}$ is updated to assess the composition of the population $\Xcal$.
      \end{enumerate*}
      The core of the update step consists in the \emph{data association}, where potential sources of observations are matched with the new observation set $Z_t$, and every resulting association is assessed. The potential sources of observations are:
      \begin{itemize}
	\item the previously-detected targets $\Xcal^{\bullet}_t$, indexed by $\Ibb^{\bullet}_{t-1}$;
	\item the yet-to-be-detected targets $\Xcal^{\circ}_t$, indexed by $\Ibb^{\circ}_{t|t-1}$;
	\item the clutter, generating the false alarms.
      \end{itemize}
      For the purpose of data association, a clutter generator is modelled for each collected observation $z \in Z_t$; as discussed in Section~\ref{subsec:filter_update_model}, it produces the observation $z$ with probability $p_{\fa, t}(z)$ (in which case $z$ is a false alarm), or it produces no observation with probability $1 - p_{\fa, t}(z)$ (in which case $z$ originates from a target).
      
      \subsubsection{Data association}   
\begin{figure*}[ht]
	  \centering
	  \includegraphics{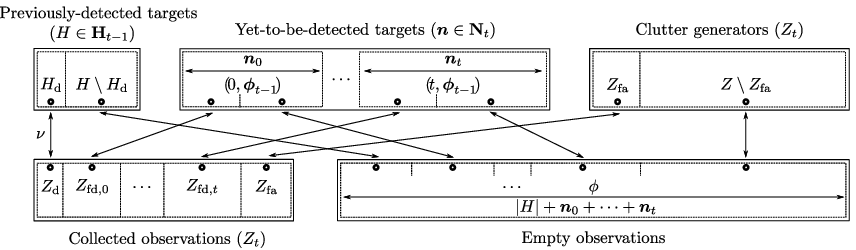}
	  \caption{Data association, for a given configuration of the population $(H, \nbs)$ and a given set of collected observations $Z_t$. \label{fig:data_association}} 
	\end{figure*}

	Let us fix a pair $(H, \nbs) \in \Hbo_{t-1} \times \Nbo_t$ corresponding to a given configuration of the population (see Eq.~\eqref{eq:prediction_total}). A data association $\hbo$ then associates the sources of observations -- the previously-detected targets described by $H$, a number of yet-to-be-detected targets described by $\nbs$, and the $|Z_t|$ clutter generators \JeremComment{~I guess we could go for a clutter generator per observation cell} -- to the collected observations $z \in Z_t$ and the empty observation $\phi$ (see Figure~\ref{fig:data_association}).
	
	Given Assumptions~\ref{mod:observation_per_target}, \ref{mod:target_per_observation}, the set of data associations $\Adm_{Z_t}(H, \nbs)$ can be defined as the set of tuples $(H_{\d}, Z_{\d}, Z_{\fd}, \nu)$ such that
	\begin{enumerate*}[label=\roman*)]
	  \item $H_{\d} \subseteq H$ contains the previously-detected targets that are detected this time step,
	  \item $Z_{\d} \subseteq Z_t$ contains the observations associated to these detected targets,
	  \item $\nu$ is a bijective function from $H_{\d}$ to $Z_{\d}$ associating detected targets with observations,
	  \item $Z_{\fd} = \{Z_{\fd, t'}\}_{t'=0}^t$ is a family of disjoint subsets of $Z_t \setminus Z_{\d}$, satisfying $|Z_{\fd, t'}| \leq \nbs_{t'}$ for any $0 \leq t' \leq t$, where $Z_{\fd, t'}$ are the observations produced by yet-to-be-detected targets that appeared at time $t'$ and are about to be detected for the first time.
	\end{enumerate*}
	
	Note that there is only one way to associate the subset of observations $Z_{\fd, t'}$ to $|Z_{\fd, t'}|$ inviduals among the $\nbs_{t'}$ composing the subpopulation of yet-to-be-detected indexed by $(t', \phibs_{t-1})$, because the latter is composed of indistinguishable targets for the purpose of estimation.
	
	Each triplet $\abo = (H, \nbs, \hbo)$, where $\hbo \in \Adm_{Z_t}(H, \nbs)$, then represents a specific \emph{association scheme} described by the scalar
	\begin{equation} 
	  v^{\abo}_t = v^{\abo}_{\d,t} \times v^{\abo}_{\md,t} \times v^{\abo}_{\fa,t}, \label{eq:association_probability}
	\end{equation}
	where
	\begin{equation} \label{eq:association_probability_detail}
	  \begin{aligned}
	    v^{\abo}_{\d,t} &= \hspace*{-5pt}\prod_{0\leq t' \leq t}\prod_{z \in Z_{\fd, t'}} \hspace*{-5pt}p^{(t',\phibs_{t-1})}_{t|t-1} (g_t(\cdot, z)) \hspace*{-2pt}\times \hspace*{-5pt}\prod_{i \in H_{\d}} p^i_{t|t-1}(g_t(\cdot, \nu(i))),
	    \\
	    v^{\abo}_{\md,t} &= \hspace*{-5pt}\prod_{0\leq t' \leq t} \hspace*{-5pt}\Big(p^{(t',\phibs_{t-1})}_{t|t-1} (g_t(\cdot,\phi))\Big)^{\nbs_{t'} - |Z_{\fd, t'}|} \hspace*{-15pt}\times \hspace*{-8pt}\prod_{i \in H \setminus H_{\d}} \hspace*{-5pt}p^i_{t|t-1}(g_t(\cdot,\phi)),
	    \\
	    v^{\abo}_{\fa,t} &= \hspace*{-8pt}\prod_{z \in Z_t \setminus Z_{\fa}} \hspace*{-5pt}(1 - p_{\fa, t}(z)) \times \prod_{z \in Z_{\fa}} p_{\fa, t}(z),
	  \end{aligned}
	\end{equation}
	where $Z_{\fa} = Z_t \setminus (Z_{\d} \cup \bigcup_{0 \leq t' \leq t} Z_{\fd, t'})$ is the subset of observations associated with clutter generators, i.e., the false alarms (see Figure~\ref{fig:data_association}).
      
      \subsubsection{Hypothesis update}
	A given association scheme ${\abo = (H, \nbs, \hbo)}$ leads to the construction of a \emph{unique} updated hypothesis $\hat{H} \subseteq \Ibb^{\bullet}_t$ of the form
	\begin{equation} \label{eq:updated_hypothesis}
	  \hat{H} = S_{\d}(H_{\d},\nu) \cup S_{\md}(H \setminus H_{\d}) \cup S_{\fd}(Z_{\fd}),
	\end{equation}
	with
	\begin{align*}
	  S_{\d}(H_{\d},\nu) & = \big\{ (t',y\concat\nu(i)) \st i = (t',y) \in H_{\d} \big\},
	  \\
	  S_{\md}(H \setminus H_{\d}) &= \big\{ (t',y \concat \phi) \st (t',y) \in H \setminus H_{\d} \big\},
	  \\
	  S_{\fd}(Z_{\fd}) & = \bigcup_{0 \leq t' \leq t} \{(t',\phibs_{t-1}\concat z) : z \in Z_{\fd, t'}\}.
	\end{align*}
	
	Also, define $\hat{\nbs} \in \Nbo_t$ as $\hat{\nbs}_{t'} = \nbs_{t'} - |Z_{\fd, t'}|$, $0 \leq t' \leq t$, such that $(\hat{H},\hat{\nbs})$ is the updated configuration under the association scheme $\abo = (H, \nbs, \hbo)$. Using Bayes' rule, the probability of existence of the updated population is found to be
	\begin{equation}
	  \wbo_t(\hat{H},\hat{\nbs}) \propto \wbo_{t|t-1}(H,\nbs)v^{\abo}_t. \label{eq:updated_hypothesis_probability}
	\end{equation}
	
	The updated hypotheses $\Hbo_t$ are constructed by considering \emph{all the possible} admissible association schemes ${\abo = (H, \nbs, \hbo)}$, where $(H, \nbs) \in \Hbo_{t-1} \times \Nbo_t$, ${\hbo \in \Adm_{Z_t}(H, \nbs)}$. Following Eq.~\eqref{eq:updated_hypothesis}, the set of updated hypotheses $\Hbo_t$ can be written as
	\begin{multline}
	  \Hbo_t = \Big\{S_{\d}(H_{\d},\nu) \cup S_{\md}(H \setminus H_{\d}) \cup S_{\fd}(Z_{\fd}) \st
	  \\
	  (H_{\d},Z_{\d},Z_{\fd},\nu) \in \Adm_{Z_t}(H, \nbs), H \in \Hbo_{t-1}, \nbs \in \Nbo_t \Big\}. \label{eq:updated_hypothesis_set}
	\end{multline}

    \subsection{Hypotheses $\Hbo_t$ and previously-detected targets $\Ibb^{\bullet}_t$} \label{subsec:filter_update_compatibility}
      We have seen in Section~\ref{subsec:filter_update_population} that the hypotheses in $\Hbo_t$ aims at representing the subsets of previously-detected targets that are pairwise-compatible, i.e., whose observation paths do not violate the ``at-most-one-measurement-per-target'' and ``at-most-one-target-per-measurement'' rules given by Assumptions~\ref{mod:observation_per_target} and \ref{mod:target_per_observation}. One may wonder whether the composition of the population of previously-detected targets $\Xcal^{\bullet}_t$ is represented by $\Hbo_t$, i.e., whether the possible configurations for $\Xcal^{\bullet}_t$ are the hypotheses $\Hbo_t$.
      
      \begin{theorem}{} \label{theo:_hypothesis_consistency}
	Define the \emph{compatibility} relation as the symmetric binary relation $\cst$ on the set of observation paths $\Ybo_t$ given by
	\begin{multline}
	  \big(\forall y, y' \in \Ybo_t\big)~ y = (z_0,\dots,z_t), y' = (z'_0,\dots,z'_t),
	  \\
	  y \cst y' \Leftrightarrow \big[\left[z_{t'} = z'_{t'}\right] \Rightarrow z_{t'} = \phi, 0 \leq t' \leq t \big], \label{eq:observation_path_compatibility}
	\end{multline}
	and the set of pairwise-compatible subsets (or \emph{consistent} subsets) of distinguishable targets as
	\begin{multline}
	  \Const(\Ibb^{\bullet}_t) = \Big\{I \subseteq \Ibb^{\bullet}_t \st \big(\forall i, i' \in I\big)~
	  \\
	  i = (t', y'), i' = (t'', y''), i \neq i' \Rightarrow y' \cst y''\Big\}. \label{eq:population_consistency}
	\end{multline}
	Then it holds that
	\begin{equation}
	  \Hbo_t = \Const(\Ibb^{\bullet}_t), \label{eq:hypothesis_consistency} 
	\end{equation}
	that is, the consistent subsets of previously-detected targets are the hypotheses.
      \end{theorem}
      The proof is given in Appendix~\ref{subsec:proof_theo_hypothesis_consistency}.

    \subsection{Output} \label{subsec:filter_update_output}
      Following Eq.~\eqref{eq:update_target} an hypothesis $H \in \Hbo_t$ is now described by the multi-target configuration measure
      \begin{equation} \label{eq:update_hypothesis}
	\mu^H_t = \sum_{i \in H} \delta [p^i_t],
      \end{equation}
      while the subpopulation of yet-to-be detected targets composed of $\nbs_{t'}$ individuals that appeared at time $t'$, $0 \leq t' \leq t$, is now described by the multi-target configuration measure
      \begin{equation} \label{eq:update_undetected}
	\mu^{\nbs}_t = \sum_{0 \leq t' \leq t} \nbs_{t'} \delta\Big[ p^{(t',\phibs_t)}_t \Big].
      \end{equation}
      The law of the stochastic population $\Cfrak_t$ maintained by the \ac{disp} filter thus becomes
      \begin{equation} \label{eq:update_total}
	P_t = \sum_{H \in \Hbo_t} \sum_{\nbs \in \Nbo_t} \wbo_t(H,\nbs) \delta\big[\mu^H_t + \mu^{\nbs}_t\big].
      \end{equation}
	
  \section{\ac{disp} filter: exploitation} \label{sec:exploitation}
    The law $P_t$ of the form \eqref{eq:population_law} propagated by the \ac{disp} filter (see Figure~\ref{fig:data_flow}) describes the stochastic population $\Cfrak_t$ defined as
    \begin{equation} \label{eq:stochastic_population_disp}
      \Cfrak_t(F) = \sum_{i \in \Ibb_t}\Nbs^i_t F(p^i_t),
    \end{equation}
    for any $F \in \Lcal^{\infty}(\Pcal(\bar{\Xbo}_t))$, where the probability mass function $\cbo_t$ of the random variable $\Nbs_t$ on $\Nbb^{\Ibb_t}$ is such that
    \begin{equation} \label{eq:probability_mass_function_disp}
      \cbo_t(\nbs) =
      \begin{dcases}
	\wbo_t(I^{\bullet}, \nbs^{\circ}), &I^{\bullet} \in \Hbo_t \textrm{~and~} \nbs_i = 1, i \in I^{\bullet},
	\\
	0, &\textrm{otherwise},
      \end{dcases}
    \end{equation}
    where $I^{\bullet} = \supp(\nbs) \cap \Ibb^{\bullet}_t$ and $\nbs^{\circ} \in \Nbo_t$ is given by $\nbs^{\circ}_{t'} = \nbs_{(t', \phibs_t)}$, $0 \leq t' \leq t$. That is, a possible configuration $\nbs \in \Nbb^{\Ibb_t}$ of the population $\Xcal$ has zero mass unless the previously-detected targets with non-zero multiplicity 
    \begin{enumerate*}[label=\roman*)]
      \item form an hypothesis (they are pairwise-compatible)
      \item have a multiplicity of one (they are distinguishable).
    \end{enumerate*}
    
    As we have seen in Section~\ref{sec:stochastic}, several tools can be designed in order to exploit a stochastic population, such as the extraction of statistical moments or merging operations. We shall see in this section how these tools can be exploited on the output of the \ac{disp} filter  for meaningful operations in a multi-target tracking scenario. 

    \subsection{\acl{map}} \label{subsec:exploitation_map}
      The operator may wish to produce a \ac{map} estimate of the population $\Xcal$ in order to display the most probable multi-target configuration from the output of the \ac{disp} filter. From the law \eqref{eq:population_law} it is straightforward to extract the most probable configuration in $\Hbo_t \times \Nbo_t$:
      \begin{equation}
	(H, \nbs)^* = \arg\max_{(H, \nbs)} \wbo_t(H, \nbs),
      \end{equation}
      that is, the previously-detected targets are most likely to be those in $H^*$, and the number of yet-to-be-detected targets that appeared at time $t'$ is most likely to be $\nbs^*_{t'}$, $0 \leq t' \leq t$.
      
      Some individual targets $i \in H^*$ might be tentative tracks that a ``cautious'' operator may not wish to display. The \emph{probability of existence} of distinguishable targets, defined in Section~\ref{subsec:exploitation_moment}, is a simple statistics assessing the likelihood of individual tracks; a simple threshold can be set on the targets' probability of existence to make sure that only the reliable ones, with a degree of confidence set by the operator, are to be displayed.
      
      The extraction of the \ac{map} state of each target to be displayed is obviously dependent on the chosen implementation technique for spatial distributions on the state space, and is left out of the scope of this paper. More details on \ac{gm} or \ac{smc} implementations can be found in \cite{Alspach_DL_1972} or \cite{Doucet_A_2001}, respectively.

    \subsection{Statistical moments} \label{subsec:exploitation_moment}
      All the quantities related to the full \eqref{eq:mean_random_conf}, \eqref{eq:var_random_conf} or collapsed \eqref{eq:mean_random_conf_collapsed}, \eqref{eq:var_random_conf_collapsed} moments of a stochastic population are produced from the statistical moments \eqref{eq:mean_pop}, \eqref{eq:var_pop} of the indexed subpopulations. In the case of the \ac{disp} filter, they are found to be
      \begin{equation} \label{eq:mean_pop_disp}
	\mbo_t(i) =
	\begin{dcases}
	  \sum_{\substack{H \in \Hbo_t \\ H \ni i}} \sum_{\nbs \in \Nbo_t} \wbo_t(H, \nbs), &i \in \Ibb^{\bullet}_t,
	  \\
	  \sum_{\substack{H \in \Hbo_t}} \sum_{\nbs \in \Nbo_t} \wbo_t(H, \nbs)\nbs_i, &i \in \Ibb^{\circ}_t,
	\end{dcases}
      \end{equation}
      and
      \begin{align} 
	&\covbo_t(i,j) \nonumber
	\\
	&=
	\begin{dcases}
	  \sum_{\substack{H \in \Hbo_t \\ H \supseteq \{i,j\}}} \sum_{\nbs \in \Nbo_t} \!\wbo_t(H, \nbs) - \!\mbo_t(i)\mbo_t(j), &i,j \in \Ibb^{\bullet}_t,
	  \\
	  \sum_{\substack{H \in \Hbo_t \\ H \ni i}} \sum_{\nbs \in \Nbo_t} \wbo_t(H, \nbs)[\nbs_j - \mbo_t(j)], &i \in \Ibb^{\bullet}_t, j \in \Ibb^{\circ}_t
	  \\
	  \sum_{\substack{H \in \Hbo_t}} \sum_{\nbs \in \Nbo_t} \wbo_t(H, \nbs)\nbs_i[\nbs_j - \mbo_t(j)], &i,j \in \Ibb^{\circ}_t.
	\end{dcases} \label{eq:cov_pop_disp}
      \end{align}
      Note in particular that the expected size $\mbo_t(i)$ of a population associated to a previously-detected target $i \in \Ibb^{\bullet}_t$, as shown in Eq.~\eqref{eq:mean_pop_disp}, is a scalar between $0$ and $1$: it is equal to
      \begin{enumerate*}[label=\roman*)]
	\item $0$ if the target $i$ does not belong to any hypothesis with non-zero mass, i.e., if the target $i$ exists \emph{almost never}, and
	\item $1$ if the target $i$ belongs to all the hypotheses with non-zero mass, i.e., if the target $i$ exists \emph{almost surely}.
      \end{enumerate*}
      For this reason, the quantity $\mbo_t(i)$ is also called the \emph{probability of existence}\footnote{Not to be confused with the probability of presence $p^i_t(\Xbo_t)$ defined in Section \ref{subsec:filter_representation_distinguishable}, assessing the probability that the target $i$ is still in the scene, provided that it exists.} of the previously-detected target $i \in \Ibb^{\bullet}_t$, and can be exploited to assess its credibility (e.g. for displaying purposes, as discussed in Section~\ref{subsec:exploitation_map}).
      
      Once the statistics \eqref{eq:mean_pop_disp}, \eqref{eq:cov_pop_disp} of the subpopulations composing the stochastic population $\Cfrak_t$ have been computed, it is straightforward to produce meaningful statistics out of the full or collapsed moments, as illustrated in Section~\ref{sec:stochastic} through Examples \ref{ex:moment_close_law} or \ref{ex:moment_close_state}. Following Example~\ref{ex:moment_close_state}, the regional statistics \cite{Delande_E_2014_4} of the population, i.e., the mean and variance of the number of targets in any subset $B \in \Bcal(\bar{\Xbo}_t)$, are given by
      \begin{align} 
	\m_t(B) &= \sum_{i \in \Ibb_t} \mbo_t(i) p^i_t(B), \label{eq:mean_random_conf_collapsed_disp}
	\\
	\var_t(B) &= \sum_{i,j \in \Ibb_t} \covbo_t(i, j) p^i_t(B)p^j_t(B), \label{eq:var_random_conf_collapsed_disp}
      \end{align}
      and provide the operator with some information on the level of target activity, with associated uncertainty, in the region $B$ (e.g. a region of strategic importance in the surveillance scene). They can be exploited, for example, in a sensor scheduling policy exploring the regions where the uncertainty is the highest \cite{Andrecki_M_2015_1_2}.

    \subsection{Information gain} \label{susbec:exploration_gain}
      In many applications, notably to establish a hierarchy among possible actions in the construction of a sensor scheduling policy, some performance metric is necessary in order to assess quantitatively a sensor action leading to the collection of some observation set $Z_t$. We proposed in \cite{Delande_E_2014_3_3} an information gain for stochastic populations, that we shall extend in this section to the \ac{disp} filter presented in this paper.
      
      We aim at quantifying the information gain from the predicted law $P_{t|t-1}$ of the form \eqref{eq:prediction_total} to the posterior law $P_t$ of the form \eqref{eq:update_total} maintained by the \ac{disp} filter, given an observation set $Z_t$ collected by the sensor system. We shall assume that the probability distributions $p \in \Pcal(\bar{\Xbo}_t)$ considered in this section admit a Radon-Nikodym derivative with respect to some reference measure $\mu \in \Mcal(\bar{\Xbo}_t)$, with $\mu(\{\psi\}) = 1$. For the sake of simplicity, the resulting probability density $\frac{\d p}{\d \mu}$ will be denoted $p$.      
      
      \subsubsection{Target gain}
	Assume some predicted track indexed by $i \in \Ibb_{t|t-1}$, $i = (t', y)$, some (possibly empty) observation $z \in \bar{Z}_t$, and the corresponding updated track indexed by ${i\concat z = (t', y\concat z) \in \Ibb_t}$ (see Section~\ref{subsec:filter_update_target}). For some mapping $F: \Pcal(\bar{\Xbo}_t) \rightarrow \Pcal(\bar{\Xbo}_t)$, to be specified later on, we define the information gain from a target indexed by $i$, updated with observation $z$, as the R\'{e}nyi divergence \cite{Renyi_A_1961}
	\begin{equation}
	  G_i^z(F) \propto \log\Big[\int_{\bar{\Xbo}_t} \!\!\!\big[F(p^i_{t|t-1})(x)\big]^{\alpha} \big[F(p^{i\concat z}_t)(x)\big]^{1 - \alpha}\mu(\d x) \Big], \label{eq:gain_individual_update}
	\end{equation}
	where the constant factor is $(\alpha - 1)^{-1}$, and $0 < \alpha < 1$ is the order of the divergence\footnote{The order of the divergence may be dependent on the track $i$, but for the sake of simplicity we will drop this dependency.}.
	
      \subsubsection{Configuration gain}
	Consider now some configuration $(H, \nbs) \in \Hbo_{t-1} \times \Nbo_t$, and some association scheme ${\abo = (H, \nbs, \hbo)}$, where $\hbo \in \Adm_{Z_t}(H, \nbs)$ (see Section~\ref{subsec:filter_update_population}). Under the association scheme $\abo$ the configuration $(H, \nbs)$ updates to $(\hat{H}, \hat{\nbs}) \in \Hbo_t \times \Nbo_t$, where the updated hypothesis $\hat{H}$ is described by Eq.~\eqref{eq:updated_hypothesis}. We can then define the information gain of the association scheme $\abo$ as
	\begin{align}
	  &G_{(H, \nbs)}^{\abo}(F) \nonumber
	  \\
	  &= \!\!\sum_{0 \leq t' \leq t} \!\Big[\!\sum_{z \in Z_{\fd, t'}} \!\!\!\!G_{(t', \phibs_{t-1})}^{z}(F) + \big[\nbs_{t'} - |Z_{\fd, t'}|\big]G_{(t', \phibs_{t-1})}^{\phi}(F) \Big] \nonumber
	  \\
	  &+ \sum_{i \in H_{\d}} G_i^{\nu(i)}(F) + \!\!\sum_{i \in H \setminus H_{\d}} G_i^{\phi}(F), \label{eq:gain_association_update}  
	\end{align}
	that is, as the sum of the individual gains \eqref{eq:gain_individual_update} from the targets in the configuration $(H, \nbs)$. The information gain for the configuration follows as
	\begin{equation}
	  G_{(H, \nbs)}(F) = \sum_{\hbo \in \Adm_{Z_t}(H, \nbs)} \!\!\!\!\!\!v^{\abo}_t G_{(H, \nbs)}^{\abo}(F), \label{eq:gain_configuration_update}  
	\end{equation}
	where the association weights $v^{\abo}_t$ are given by Eq.~\eqref{eq:association_probability}.
	
      \subsubsection{Population gain}
	The expected information gain for the stochastic population with law $P_{t|t-1}$ of the form \eqref{eq:prediction_total}, given the collected observations $Z_t$, is thus given by
	\begin{subequations} \label{eq:gain_population_update}
	  \begin{align}
	     G_t(F) &= \Exp\Big[G_{(H, \nbs)}(F)\Big] \label{eq:gain_population_update_1}
	     \\
	     &= \sum_{H \in \Hbo_{t-1}} \sum_{\nbs \in \Nbo_t} \wbo_{t|t-1}(H, \nbs) G_{(H, \nbs)}(F), \label{eq:gain_population_update_2}
	  \end{align}
	\end{subequations}
	where the expectation in \eqref{eq:gain_population_update_1} is taken \ac{wrt} the configurations in $\Hbo_{t-1} \times \Nbo_t$.
	
      \subsubsection{Evaluation of the population gain}
	We can now shape the mapping $F$ in order to focus the evaluation of the information gain \eqref{eq:gain_population_update} towards specific targets and/or specific regions of the state space of interest to the operator \cite{Delande_E_2014_3_3}.
	
	Consider the probability distribution $p_{\psi} \in \Pcal(\bar{\Xbo}_t)$ defined as $p_{\psi}(\{\psi\}) = 1$, i.e., all the mass is concentrated on the empty target state $\psi$. For any subset $I \subseteq \Ibb_{t|t-1}$, consider the mapping $F^I$ defined as
	\begin{align}
	  F^I(p^i_{t|t-1}) &= 
	  \begin{dcases}
	   p^i_{t|t-1}, &i \in I
	   \\
	   p_{\psi}, &\textrm{otherwise},
	  \end{dcases}
	  \\
	  F^I(p^{i \concat z}_t) &= 
	  \begin{dcases}
	   p^{i\concat z}_t, \hspace{7pt}&i \in I,
	   \\
	   p_{\psi}, &\textrm{otherwise}.
	  \end{dcases}
	\end{align}
	The individual gain \eqref{eq:gain_individual_update} then becomes
	\begin{align} \label{eq:gain_individual_update_target_specific}
	  &G_i^z(F^I) \nonumber
	  \\
	  &\propto
	  \begin{dcases}
	    \log\Big[\int_{\bar{\Xbo}_t}\!\!\big[p^i_{t|t-1}(x)\big]^{\alpha} \!\big[p^{i\concat z}_t(x)\big]^{1 - \alpha} \mu(\d x) \Big], &i \in I
	    \\
	    0, &\textrm{otherwise}.
	  \end{dcases}
	\end{align} 
	That is, the scalar $G_t(F^I)$ quantifies the information gain, provided by the collected observations $Z_t$, regarding the targets $i \in I$. In particular, the scalar $G_t(F^{\Ibb_{t|t-1}})$ quantifies the information gain regarding the whole population.
	
	For any subset $B \subseteq \Bcal(\Xbo_t)$, consider now the mapping $f_B: \bar{\Xbo}_t \rightarrow \bar{\Xbo}_t$ defined as
	\begin{align}
	  f_B(x) &= 
	  \begin{dcases}
	   x, &x \in B
	   \\
	   \psi, &\textrm{otherwise},
	  \end{dcases}
	\end{align}
	and define $F_B(p)$ as the image of the measure $p \in \Pcal(\bar{\Xbo}_t)$ under $f_B$ (see Section 3.6. in \cite{Bogachev_VI_2007}), or pushforward measure, defined by
	\begin{equation}
	  (F_B(p))(\d x) = p(f_B^{-1}(\d x)).
	\end{equation}
	
	The individual gain \eqref{eq:gain_individual_update} then becomes
	\begin{multline}
	  G_i^z(F_B) \propto \log\Big[\int_B \big[p^i_{t|t-1}(x)\big]^{\alpha}
	  \big[ p^{i \concat z}_t(x)\big]^{1 - \alpha}\mu(\d x)
	  \\
	  + \big[p^i_{t|t-1}(\bar{\Xbo}_t \setminus B)\big]^{\alpha}\big[p^{i\concat z}_t(\bar{\Xbo}_t \setminus B)\big]^{1 - \alpha} \Big]. \label{eq:gain_individual_update_region_specific}
	\end{multline}
	We see that the gain is zero \ac{iff} the (possibly empty) observation $z$ carried no additional information on the target $i$ regarding:
	\begin{itemize}
	  \item its localization in $B$, since $p^i_{t|t-1} = p^{i\concat z}_t$ on $B$, and
	  \item its presence in $B$, since $p^i_{t|t-1}(\bar{\Xbo}_t \setminus B) = p^{i\concat z}_t(\bar{\Xbo}_t \setminus B)$.
	\end{itemize}
	That is, the scalar $G_t(F_B)$ quantifies the information gain, provided by the collected observations $Z_t$, within the region $B$. In particular, the scalar $G_t(F_{\Xbo_t})$ quantifies the information gain in the whole state space.
	
	In addition, the scalar $G_t(F_B \circ F^I)$ quantifies the information gain, provided by the collected observations $Z_t$, regarding the targets $i \in I$ \emph{and} within the region $B$.
	
      \subsubsection{Knowledge-based sensor policy}
	The population gain \eqref{eq:gain_population_update} extends the results in \cite{Delande_E_2014_3_3}, exploited for an approximate version of the \ac{disp} filter (briefly discussed in Section~\ref{subsec:exploitation_merging}), to the general version of the \ac{disp} filter introduced in this paper. A similar approach to \cite{Delande_E_2014_3_3} could then be applied to produce a closed-loop sensor scheduling policy focusing on sensing actions yielding the highest expected information gain, though this is left out of the scope of this paper.
	
	Note that considering subsets of the form $I \subseteq \Ibb^{\bullet}_{t|t-1}$, in the expression of the population gain \eqref{eq:gain_individual_update_target_specific}, would favour sensor actions \emph{exploiting} previously-detected targets; conversely, considering subsets of the form $I \subseteq \Ibb^{\circ}_{t|t-1}$ would favour sensor actions \emph{exploring} the state space for the search of yet-to-be-detected targets.
%
	
    \subsection{Merging operations} \label{subsec:exploitation_merging}
      Merging operations are important for practical implementations of multi-target tracking algorithms, for they allow to mitigate the loss of information whenever approximated versions of the algorithms are necessary to curtail computational costs. We shall highlight in this section a specific merging operation leading to a principled approximation of the \ac{disp} filter whose complexity is significantly reduced\footnote{Another principled approximation, based on additional modelling assumptions and leading to a much simpler structure, yields the filter for \ac{hisp} \cite[chap. 4]{Houssineau_J_2015_1_2}.}.
      
      In its most general form \eqref{eq:population_law}, the \ac{disp} filter maintains specific representations for populations of yet-to-be-detected targets that appeared at different times. If the prior information on appearing targets \eqref{eq:appearing_law} varies little across time, however, the probability distributions of the undetected targets ${\{p_{t|t-1}^i \st i \in \Ibb^{\circ}_{t-1}\}}$ are likely to be close to each other, and close to the probability distribution describing the appearing targets $p^{i_{\a}}_{t|t-1}$. In this context, one may wish to merge \emph{all} the subpopulations of yet-to-be-detected targets indexed in $\Ibb^{\circ}_{t-1}$, alongside the population of appearing targets indexed by $i_{\a}$, into a single population describing all the yet-to-be-detected targets \emph{so far}, i.e., \emph{regardless of their time of arrival}.
      
      Suppose that these merging operations have been conducted until the present time $t \geq 0$. Since the appearing targets were merged in past times regardless of their time of arrival, the targets indexed in $\Ibb_{t-1}$ are characterized with their observation path, i.e.
      \begin{align}
	\Ibb^{\bullet}_{t-1} &= \Ybo_{t-1}, \label{eq:simplified_label_distinguishable}
	\\
	\Ibb^{\circ}_{t-1} &= \{\phibs_{t-1}\}. \label{eq:simplified_label_indistinguishable}
      \end{align}
      The multi-target configuration \eqref{eq:configuration_indistinguishable} for yet-to-be-detected targets at time $t-1$ then takes the simpler form
      \begin{equation} \label{eq:configuration_indistinguishable_approx}
	\mu^{(n)}_{t-1} = n\delta\Big[p^{\phibs_{t-1}}_{t-1}\Big],
      \end{equation}
      such that the posterior information \eqref{eq:population_law} at time $t-1$ takes the simpler form
      \begin{equation} \label{eq:population_law_approx}
	P_{t-1} = \sum_{H \in \Hbo_{t-1}} \sum_{n \in \Nbb} \wbo_{t-1}(H, n) \delta\big[\mu^H_{t-1} + \mu^{(n)}_{t-1}\big].
      \end{equation}
      Using Eq.~\eqref{eq:merged_population_cardinality}, the merging of the yet-to-be-detected targets indexed by $\phibs_{t-1}$ and the appearing targets indexed by $i_{\a}$ transforms the prediction of the probability mass function \eqref{eq:prediction_mass_function} to the simpler form
      \begin{multline}
	\wbo_{t|t-1}(H, n) =
	\\
	\begin{dcases}
	  \sum_{0 \leq n' \leq n} \wbo_{t-1}(H, n') \rho^{\a}_t(n - n'), &t > 0
	  \\
	  \rho^{\a}_0(n), &t = 0,
	\end{dcases} \label{eq:prediction_mass_function_approx}
      \end{multline}
      for any $H \in \Hbo_{t-1}$ and any $n \in \Nbb$.
      
      Once the probability distributions $\{p_{t-1}^i \st i \in \Ibb_{t-1}\}$ have been transformed to their predicted values $\{p_{t|t-1}^i \st i \in \Ibb_{t-1}\}$ according to Eq.~\eqref{eq:prediction_track}, the laws $p_{t|t-1}^{\phibs_{t-1}}$ and $p_{t|t-1}^{i_{\a}}$ can be mixed according to Eq.~\eqref{eq:merged_population_law_unique}. The predicted law \eqref{eq:prediction_total} then takes the simpler form 
      \begin{equation} \label{eq:prediction_total_approx}
	P_{t|t-1} = \sum_{H \in \Hbo_{t-1}} \sum_{n \in \Nbb} \wbo_{t|t-1}(H,n) \delta\big[\mu^H_{t|t-1} + \mu^{(n)}_{t|t-1}\big].
      \end{equation}
    
      The main simplification occurs in the data association of the update step (see Section~\ref{subsec:filter_update_population}). For a given configuration $(H, n) \in \Hbo_{t-1} \times \Nbb$, the set of data associations $\Adm_{Z_t}(H, n)$ can be defined as the set of tuples $(H_{\d}, Z_{\d}, Z_{\fd}, \nu)$ such that
      \begin{enumerate*}[label=\roman*)]
	\item $H_{\d} \subseteq H$ contains the previously-detected targets that are detected this time step,
	\item $Z_{\d} \subseteq Z_t$ contains the observations associated to these detected targets,
	\item $\nu$ is a bijective function from $H_{\d}$ to $Z_{\d}$ associating detected targets with observations,
	\item $Z_{\fd}$ is a subset of $Z_t \setminus Z_{\d}$ satisfying $|Z_{\fd}| \leq n$, containing the observations produced by the yet-to-be-detected targets that are about to be detected for the first time.
      \end{enumerate*}
      
\begin{figure*}[ht]
	\centering
	\includegraphics{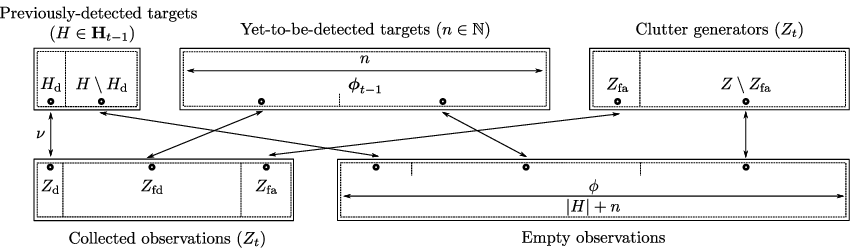}
	\caption{Data association (approximated version), for a given configuration of the population $(H, n)$ and a given set of collected observations $Z_t$. \label{fig:data_association_approx}} 
      \end{figure*}

      The subset $Z_{\fd}$ of observations allocated to first detections does not have to be partitioned among the subpopulations of yet-to-be-detected targets appearing at different times, resulting in a simplified data association mechanism (see the simplified structure illustrated in Figure~\ref{fig:data_association_approx}, compared to the general structure illustrated in Figure~\ref{fig:data_association}). The number of hypotheses created in the update step is thus reduced, and the number of subpopulations of yet-to-be-detected targets in the stochastic population $\Cfrak_t$ remains fixed at one, rather than growing of one at every time step. However, as seen in Eqs~\eqref{eq:simplified_label_distinguishable} and \eqref{eq:simplified_label_indistinguishable}, the information about the time of arrival of the targets has been lost in the merging operation and this approximated \ac{disp} filter cannot estimate the period of time for which a given target has been in the surveillance scene prior to its first detection.

      A further approximated version of the \ac{disp} filter has been proposed, in which appearing targets are assumed to be detected at their time of arrival \cite{Delande_E_2015_1_2}. In this case, since targets are assumed detected upon entering the scene, no information is propagated across time on yet-to-be-detected targets; in other words, this approximated version of the \ac{disp} loses the ability to estimate the number and states of the targets that have not been detected yet. This approximation has been recently exploited for a multi-target tracking problem in the context of space situational awareness \cite{Delande_E_2016_1_1}.\newpage
	
  \section{Conclusion} \label{sec:conclusion}
    In this paper, we present an estimation framework for stochastic populations as an original approach for multi-target estimation problems, and the derivation of a detection and tracking algorithm adapted to challenging scenarios, the \ac{disp} filter, maintaining information on the targets that have been previously detected by the sensor system observing the scene, as well as those that have not been detected yet. We show that the fully probabilistic nature of the framework can be exploited to produce meaningful quantities describing the output of the \ac{disp} filter, including statistical moments on the population of targets or information-theoretical gain functions assessing the utility of a set of observations collected from the sensor system. This paper focussed on the methodological developments of multi-object filters; forthcoming works will focus on the illustration of the novel concepts introduced in this paper.
    
  \appendices
  \section{Multi-target configuration measures and joint probability measures} \label{sec:equivalence_mu_P}
    It can be shown \cite[chap. 2]{Houssineau_J_2015_1_2} than any multi-target configuration measure of the form \eqref{eq:configuration_with_integer_measures} induces an equivalent joint probability measure defined as follows. First, we introduce the set $E^{\times}$ as
    \begin{equation}
      E^{\times} = \bigcup_{k \geq 1} E^k,
    \end{equation}
    and the set $\Nbb_{I}$ as
    \begin{equation} \label{eq:N_I}
      \Nbb_{\Ibb} = \{(I, \nbs) \st I \subseteq \Ibb, \nbs \in \Nbb^{I}\}.
    \end{equation}
    For a given pair $(I, \nbs) \in \Nbb_{\Ibb}$, we then introduce the sets
    \begin{align}
      E_I^{(\nbs)} &= \{X \in (E^{\times})^I \st (\forall i \in I)~ X_i \in E^{\nbs_i}\},
      \\
      E^{\times}_{\Ibb} &= \bigcup_{(I, \nbs) \in \Nbb_{\Ibb}} E_I^{(\nbs)}.
    \end{align}
    Then, for a given multiplicity $\nbs \in \Nbb^{\Ibb}$ and denoting by $I = \supp(\nbs)$ its support, the multi-configuration $\mu_{\nbs}$ induces the joint probability measure $P(\cdot|I, \nbs) \in \Pcal(E^{\times}_{\Ibb})$ defined as
    \begin{equation}
      \big(\forall \Bbs \in \Bcal(E_I^{(\nbs)})\big)~~ P(\Bbs|I, \nbs) = \prod_{i \in I} p_i^{\times \nbs_i}(\Bbs_i).
    \end{equation}
    where, for any $n \in \Nbb$ and any $p \in \Pcal(E)$, $p^{\times n} \in \Pcal(E^{n})$ is defined as
    \begin{equation}
      \big(\forall B = (B_1, \ldots, B_{n}) \in \Bcal(E^n)\big)~~ p^{\times n}(B) = \prod_{i = 1}^n p(B_i).
    \end{equation}
    Note that all the mass of the joint probability measure $P(\cdot|I, \nbs)$ is concentrated in $E_I^{(\nbs)}$, i.e, in the elements of $E^{\times}_{\Ibb}$ agreeing with the multiplicity $\nbs$.
    
    \begin{example}
      Assume $\Ibb = \{\a, \b, \c\}$, the target laws $p_{\a}$, $p_{\b}, p_{\c}$, the subset $I = \{\a, \b\}$, and the multiplicity $\nbs \in \Nbb^{\Ibb}$ such that $\nbs_{\a} = 1$, $\nbs_{\b} = 2$, $\nbs_{\c} = 0$. The multi-target configuration $\mu_{\nbs}$ is then
      \begin{equation}
	\mu_{\nbs} = \delta[p_{\a}] + 2\delta_[p_{\b}].
      \end{equation}
      Consider the elements $\Bbs, \Bbs' \in \Bcal(E^{\times}_{\Ibb})$ with $\Bbs_{\a} = A$, $\Bbs_{\b} = (A, B)$, $\Bbs'_{\a} = A$, $\Bbs'_{\c} = B$. The probability that there is one target indexed with $\a$ lying within $A$, two targets indexed with $\b$, one lying within $A$ and one within $B$, and no target indexed with $\c$, is then
      \begin{equation}
	P(\Bbs|I, \nbs) = p_{\a}(A)p_{\b}(A)p_{\b}(B),
      \end{equation}
      and the probability that there is one target indexed with $\a$ lying within $A$, one target indexed with $\c$ lying within $B$, and no target indexed with $\b$, is
      \begin{equation}
	P(\Bbs'|I, \nbs) = 0,
      \end{equation}
      since $\Bbs' \notin \Bcal(E^{(\nbs)}_{I})$.
    \end{example}
%
    
  \section{Proofs} \label{sec:proof}
    \subsection{Property \ref{prop:prediction_probability_presence}} \label{subsec:proof_prop_prediction_probability_presence}
      \begin{proof}
	Let $i \in \Ibb_{t-1}$, and let $f \in \Lcal^{\infty}(\bar{\Xbo}_t)$ be an arbitrary function. Substituting the expression of the transformation \eqref{eq:gamma_transform} into the definition of the predicted distribution \eqref{eq:prediction_track} yields
	\begin{subequations} \label{eq:proof_expression_predicted_probability_distribution}
	  \begin{align}
	    p^i_{t|t-1}(f) &= p^i_{t-1}(m_{t-1, t}(\cdot, f)) \label{eq:proof_expression_predicted_probability_distribution_A}
	    \\
	    &= \int_{\bar{\Xbo}_{t-1}} \left[\int_{\bar{\Xbo}_t} \!\!\!f(x')m_{t-1, t}(x, \d x') \right] p^i_{t-1}(\d x).\label{eq:proof_expression_predicted_probability_distribution_end}
	  \end{align}
	\end{subequations}
	The probability of presence of the predicted track is then given by selecting $f = 1_{\Xbo_t}$ in Eq.~\eqref{eq:proof_expression_predicted_probability_distribution_end}, i.e.
	\begin{equation}
	  p^i_{t|t-1}(\Xbo_t) = \int_{\bar{\Xbo}_{t-1}} \left[\int_{\Xbo_t} m_{t-1, t}(x, \d x') \right] p^i_{t-1}(\d x). \label{eq:proof_expression_predicted_probability_presence_beg}
	\end{equation}
	Now, considering the expression of the Markov transition kernel \eqref{eq:prediction_transition}, Eq.~\eqref{eq:proof_expression_predicted_probability_presence_beg} simplifies to
	\begin{subequations} \label{eq:proof_expression_predicted_presence}
	  \begin{align}
	    &p^i_{t|t-1}(\Xbo_t)\nonumber
	    \\
	    &= \int_{\Xbo_{t-1}} \!\!\left[\int_{\Xbo_t} \!\!\!p_{\s,t}(x)\hat{m}_{t-1, t}(x, \d x') \right] p^i_{t-1}(\d x) \label{eq:proof_expression_predicted_probability_presence_A}
	    \\
	    &= \int_{\Xbo_{t-1}} \!\!\!\!p_{\s,t}(x) \underbrace{\left[\int_{\Xbo_t} \!\!\hat{m}_{t-1, t}(x, \d x') \right]}_{= 1} p^i_{t-1}(\d x) \label{eq:proof_expression_predicted_probability_presence_B}
	    \\
	    &= p^i_{t-1}(\Xbo_{t-1}) + \int_{\Xbo_{t-1}} \!\!\!\!\![p_{\s,t}(x) - 1]p^i_{t-1}(\d x). \label{eq:proof_expression_predicted_probability_presence_D}
	  \end{align}
	\end{subequations}
      \end{proof}
      
    \vspace*{-10pt}
    \subsection{Property \ref{prop:update_probability_presence}} \label{subsec:proof_prop_update_probability_presence}
      \begin{proof}
	Let $i = (t', y) \in \Ibb_{t|t-1}$, $z \in \bar{Z}_t$, $i\concat z = (t', y\concat z) \in \Ibb_t$, and let $f \in \Lcal^{\infty}(\bar{\Xbo}_t)$ be an arbitrary function. The probability distribution of the predicted track $i$ can be decomposed as follows:
	\begin{equation} \label{eq:proof_decomposition_probability_distribution}
	  p^i_{t|t-1}(f) = p^i_{t|t-1} (1_{\{\psi\}}f) + p^i_{t|t-1} (1_{\Xbo_t}f).
	\end{equation}
	Now, substituting the expression of the Boltzmann-Gibbs transformation \eqref{eq:BoltzmannGibbs_detail} into the definition of the posterior distribution \eqref{eq:update_target} yields
	\begin{subequations} \label{eq:proof_expression_updated_probability_distribution}
	  \begin{align}
	    p^{i\concat z}_t(f) &= \frac{p^i_{t|t-1}(g_t(\cdot, z)f)}{p^i_{t|t-1}(g_t(\cdot, z))} \label{eq:proof_expression_updated_probability_distribution_A}
	    \\
	    &= \frac{p^i_{t|t-1}(g_t(\cdot, z)1_{\{\psi\}}f) + p^i_{t|t-1} (g_t(\cdot, z)1_{\Xbo_t}f)}{p^i_{t|t-1}(g_t(\cdot, z)1_{\{\psi\}}) + p^i_{t|t-1} (g_t(\cdot, z)1_{\Xbo_t})} \label{eq:proof_expression_updated_probability_distribution_B}
	    \\      
	    &= \frac{p^i_{t|t-1}(g_t(\cdot, z)1_{\{\psi\}}f) + p^i_{t|t-1} (g_t(\cdot, z)1_{\Xbo_t}f)}{(1 - p^i_{t|t-1}(\Xbo_t))g_t(\psi, z) + p^i_{t|t-1}(g_t(\cdot, z) 1_{\Xbo_t})}, \label{eq:proof_expression_updated_probability_distribution_end}
	  \end{align}
	\end{subequations}
	where we substitute Eq.~\eqref{eq:proof_decomposition_probability_distribution} into Eq.~\eqref{eq:proof_expression_updated_probability_distribution_A} to yield the result \eqref{eq:proof_expression_updated_probability_distribution_B}. The probability of presence of the updated track $i\concat z$ is then given by selecting $f = 1_{\Xbo_t}$ in Eq.~\eqref{eq:proof_expression_updated_probability_distribution_end}, i.e.,
	\begin{equation} \label{eq:proof_update_probability_presence}
	  p^{i\concat z}_t(\Xbo_t) = \frac{p^i_{t|t-1} (g_t(\cdot, z)1_{\Xbo_t})}{(1 - p^i_{t|t-1}(\Xbo_t))g_t(\psi, z) + p^i_{t|t-1}(g_t(\cdot, z) 1_{\Xbo_t})}.
	\end{equation}
	Suppose that $z \in Z_t$. Then, exploiting the stochastic description of the sensor system \eqref{eq:update_likelihood} gives $g_t(\psi, z) = 0$, and therefore the probability of presence \eqref{eq:proof_update_probability_presence} becomes
	\begin{equation} \label{eq:proof_update_probability_presence_detection}
	  p^{i\concat z}_t(\Xbo_t) = 1.
	\end{equation}
	Suppose that $z = \phi$. Then, exploiting the stochastic description of the sensor system \eqref{eq:update_likelihood} gives $g_t(\psi, z) = 1$, and therefore the probability of presence \eqref{eq:proof_update_probability_presence} becomes
	\begin{equation} \label{eq:proof_update_probability_presence_miss_detection}
	  p^{i\concat z}_t(\Xbo_t) = \frac{p^i_{t|t-1}(g_t(\cdot, \phi)1_{\Xbo_t})}{1 - p^i_{t|t-1}(\Xbo_t) + p^i_{t|t-1}(g_t(\cdot, \phi)1_{\Xbo_t})}.
	\end{equation}
      \end{proof}
    \subsection{Theorem \ref{theo:_hypothesis_consistency}} \label{subsec:proof_theo_hypothesis_consistency}
      \begin{proof}
	The proof follows an inductive approach. By definition of the original time, no observation were produced before $t = 0$; therefore, the initial set of observation paths $\Ybo_{-1}$ is empty. It follows from the definition of consistent populations \eqref{eq:population_consistency} that $\Const(I^{\bullet}_{-1})$ is reduced to the ``empty'' hypothesis, i.e., $\Const(I^{\bullet}_{-1}) = \{\emptyset\}$. Since, by assumption, the set of hypotheses is initialized with the ``empty'' hypothesis, it follows that the base case
	\begin{equation} \label{eq:proof_hypothesis_consistency_base_case}
	  \Hbo_{-1} = \Const(I^{\bullet}_{-1})
	\end{equation}
	is true. Let us now suppose that the case is true at some rank $t - 1 \geq -1$, i.e., $\Hbo_{t-1} = \Const(\Ibb^{\bullet}_{t-1})$, and let us prove that it is true at rank $t$.
	
	\subsubsection{$\Const(\Ibb^{\bullet}_t) \subseteq \Hbo_t$}
	  Let $I \in \Const(\Ibb^{\bullet}_t)$, and let $I_{\p}$ be the set of previously-detected parent targets from $I$, i.e., the set
	  \begin{equation}
	    I_{\p} = \{(t', y) \in \Ibb^{\bullet}_{t-1} \st (t', y \concat z) \in I, z \in \bar{Z}_t\}.
	  \end{equation}
	  By construction $I_{\p} \subseteq \Ibb^{\bullet}_{t-1}$, and if $|I_{\p}| \leq 1$ we have immediately $I_{\p} \in \Const(\Ibb^{\bullet}_{t-1})$. Let us suppose that ${|I_{\p}| > 2}$, then let $(t', y'), (t'', y'') \in I_{\p}$ with $(t', y') \neq (t'', y'')$, where ${y' = (z'_{0},\dots,z'_{t-1})}$ and $y'' = (z''_{0},\dots,z''_{t-1})$. Since $I \in \Const(\Ibb^{\bullet}_t)$, there exist $z'_t, z''_t \in \bar{Z}_t$ such that $(t', y' \concat z'), (t'', y''\concat z'') \in \Ibb^{\bullet}_t$ and $y' \concat z' \cst y''\concat z''$. Then by definition of the track compatibility \eqref{eq:observation_path_compatibility}:
	  \begin{equation} \label{eq:proof_observation_path_compatibility}
	    \left[z'_{t'''} = z''_{t'''}, 0 \leq t''' \leq t\right] \Rightarrow z'_{t'''} = \phi,
	  \end{equation}
	  which implies that $y' \cst y''$. Thus, using the definition of consistent populations \eqref{eq:population_consistency}, $I_{\p} \in \Const(\Ibb^{\bullet}_{t-1})$. Using the case at rank $t - 1$, we conclude that $I_{\p} \in \Hbo_{t-1}$. Now, let us define:
	  \begin{itemize}
	    \item $I_{\d} = \{(t', y') \in I_{\p} \st (t', y' \concat z) \in I, z \in Z_t\}$;
	    \item $Z_{\d} = \{z \in Z_t \st (t', y' \concat z) \in I, (t', y') \in I_{\d}\}$;
	    \item $\nu$ is the bijective function from $I_{\d}$ to $Z_{\d}$ such that $(t', y'\concat \nu(i)) \in I$, $(t', y') \in I_{\d}$;
	    \item $Z_{\fd, t'} = \{z \in Z_t \st (t', \phibs_{t-1} \concat z) \in I\}$, $0 \leq t' \leq t$;
	    \item $\nbs_{t'} = |Z_{\fd, t'}|$, $0 \leq t' \leq t$.
	  \end{itemize}
	  Then, since $I_{\p} \in \Hbo_{t-1}$ and considering the data association
	  \begin{equation} \label{eq:proof_data_association}
	    \hbo = (I_{\d}, Z_{\d}, Z_{\fd}, \nbs) \in \Adm_{Z_t}(I_{\p}, \nbs),
	  \end{equation}
	  then the association scheme $\abo = (I_{\p}, \nbs, \hbo)$ leads to the construction of hypothesis $I$; in other words, $I \in \Hbo_t$.

	\subsubsection{$\Hbo_t \subseteq \Const(\Ibb^{\bullet}_t)$}
	  Let $H \in \Hbo_t$, and let $\abo = (H_{\p}, \nbs, \hbo)$, where $(H_{\p}, \nbs) \in \Hbo_{t-1} \times \Nbo_t$, $\hbo \in \Adm_{Z_t}(H_{\p}, \nbs)$, be the association scheme that produced hypothesis $H$ (see Eq.~\eqref{eq:updated_hypothesis}), with ${\hbo = (H_{\d}, Z_{\d}, Z_{\fd}, \nu)}$. By construction $H \subseteq \Ibb^{\bullet}_t$, and if $|H| \leq 1$ we have immediately $H \in \Const(\Ibb^{\bullet}_t)$.  Let us suppose that $|H| > 2$, and let $(t', y'), (t'', y'') \in H$ with ${(t', y') \neq (t'', y'')}$.
	  
	  1. Assume $y' = \phibs_{t-1}\concat z'$ and $y' = \phibs_{t-1}\concat z''$. If $z' = z''$ then $y = y'$, and also $z'' \in Z_{\fd, t'}$ and thus $t'' = t'$, which contradicts the assumption $(t', y') \neq (t'', y'')$. Thus $z' \neq z''$, and therefore $y' \cst y''$.
	  
	  2. Assume $y' = \phibs_{t-1}\concat z'$ and $y'' = y''_{\p}\concat z''$, where ${(t'', y''_{\p}) \in H_{\p}}$. Then either $z'' = \phi$ or $z'' \in Z_{\d}$, in both options $z' \neq z''$ since $z' \in  Z_{\fd, t'}$, and therefore $y' \cst y''$. The same reasoning applies if $y' = y'_{\p}\concat z'$ and $y'' = \phibs_{t-1}\concat z''$.
	  
	  3. Assume $y' = y'_{\p}\concat z'$ and $y'' = y''_{\p}\concat z''$, with $(t', y'_{\p}), (t'', y''_{\p}) \in H_{\p}$. Let us prove that ${(t', y'_{\p}) \neq (t'', y''_{\p})}$; for that, let us suppose that $(t', y'_{\p}) = (t'', y''_{\p})$. If ${(t', y'_{\p}) \in H_{\d}}$, then $z' = \nu((t', y'_{\p})) = \nu((t'', y''_{\p})) = z''$; if $(t', y'_{\p}) \in H_{\p} \setminus H_{\d}$, then $z = \phi = z'$; in both options $y' = y''$, and since $(t', y'_{\p}) = (t'', y''_{\p})$ it follows that $(t', y') = (t'', y'')$, which contradicts the assumption $(t', y') \neq (t'', y'')$. We thus have $(t', y'_{\p}) \neq (t'', y''_{\p})$. Using the case at rank $t - 1$ we have also $\{(t', y'_{\p}), (t'', y''_{\p})\} \subseteq H_{\p} \in \Hbo_{t-1} = \Const(Y^{\bullet}_{t-1})$ and the definition of the track compatibility \eqref{eq:observation_path_compatibility} therefore yields
	  \begin{equation} \label{eq:proof_observation_path_compatibility_parent}
	    y'_{\p} \cst y''_{\p}.
	  \end{equation}
	  Let us now suppose that $z' = z''$. If $z' \neq \phi$, then ${(t', y'_{\p}) = \nu^{-1}(z') = \nu^{-1}(z'') = (t'', y''_{\p})}$, which contradicts the fact that $(t', y'_{\p}) \neq (t'', y''_{\p})$. Therefore
	  \begin{equation} \label{eq:proof_new_observation}
	    \left[z' = z''\right] \Rightarrow z' = \phi,
	  \end{equation}
	  and from Eqs~\eqref{eq:proof_observation_path_compatibility_parent} and \eqref{eq:proof_new_observation} we conclude that $y \cst y'$.
	  
	  In all three options above we have $y \cst y'$, and therefore $H \in \Const(\Ibb^{\bullet}_t)$.
	  
	\subsubsection{$\Hbo_t = \Const(\Ibb^{\bullet}_t)$}
	  Since $\Const(\Ibb^{\bullet}_t) \subseteq \Hbo_t$ and ${\Hbo_t \subseteq \Const(\Ibb^{\bullet}_t)}$ it follows that $\Hbo_t = \Const(\Ibb^{\bullet}_t)$, which proves the case at rank $t$.
      \end{proof}   

  \bibliography{Bibliography/Bibliography.bib}
  \bibliographystyle{IEEEtran.bst}
\end{document}